\PassOptionsToPackage{english}{babel}
\documentclass[aps,pra,reprint,superscriptaddress,letterpaper,twocolumn,longbibliography,floatfix]{revtex4-2}
\usepackage{graphicx}
\usepackage{amssymb}
\usepackage[english,ngerman]{babel}

\usepackage[usenames, dvipsnames, x11names]{xcolor}

\usepackage[colorlinks]{hyperref}
\hypersetup{
 colorlinks=true,
 citecolor=red,
 linkcolor=blue,
 urlcolor=cyan}

\usepackage[english]{babel}
\usepackage[utf8]{inputenc}
\usepackage{blindtext}
\usepackage{amsmath}
\usepackage[T1]{fontenc}
\usepackage{dsfont}
\usepackage{esint}
\usepackage{mathtools}
\usepackage{array}
\DeclareUnicodeCharacter{2009}{\,}

\newcommand{\nbar}{\bar{n}}

\newcommand{\nbard}{\bar{n}_{\rm d}}
\newcommand{\nbarsq}{\bar{n}_{\rm s}}
\newcommand{\nbartot}{\bar{n}_{\rm inc}}

\newcommand{\nbardi}{\bar{n}_{1}}
\newcommand{\nbarsqi}{\bar{n}_{2}}
\newcommand{\nbartoti}{\bar{n}_{\rm cav}}

\newcommand{\as}{\alpha_{\rm s}}
\newcommand{\gff}{\gamma_{\rm ff}}
\newcommand{\detr}{\beta_{\rm s}}
\newcommand{\LamImp}{\Lambda_{\rm imp}}



\begin{document}

\selectlanguage{english}
\title{Surpassing spectator qubits with photonic modes and continuous measurement for Heisenberg-limited noise mitigation}
\date{\today}
\author{Andrew Lingenfelter$^{1,2*}$ and Aashish A. Clerk}
\affiliation{Pritzker School of Molecular Engineering, University of Chicago, Chicago, IL 60637, USA \\
$^2$Department of Physics, University of Chicago, Chicago, IL 60637, USA\\
${}^*${\rm lingenfelter@uchicago.edu}}

\begin{abstract}
Noise is an ever-present challenge to the creation and preservation of fragile quantum states. 
Recent work suggests that spatial noise correlations can be harnessed as a resource for noise mitigation via the use of spectator qubits to measure environmental noise.
In this work we generalize this concept from spectator qubits to a spectator mode: a photonic mode which continuously measures spatially correlated classical dephasing noise and applies a continuous correction drive to frequency-tunable data qubits. 
Our analysis shows that by using many photon states, spectator modes can surpass many of the quantum measurement constraints that limit spectator qubit approaches.  
We also find that long-time data qubit dephasing can be arbitrarily suppressed, even for white noise dephasing.
Further, using a squeezing (parametric) drive, the error in the spectator mode approach can exhibit Heisenberg-limited scaling in the number of photons used.
We also show that spectator mode noise mitigation can be implemented completely autonomously using engineered dissipation.  In this case no explicit measurement or processing of a classical measurement record is needed.  Our work establishes spectator modes as a potentially powerful alternative to spectator qubits for noise mitigation. 
\end{abstract}

\maketitle


\section{Introduction} \label{sec:introduction}

The protection of quantum states against decoherence due to noise is a fundamental challenge to robust quantum information processing. 
With recent progress in the scale-up of quantum hardware, efficient noise mitigation for many qubit systems is increasingly necessary \cite{preskill_Quantum_2018,arute_Quantum_2019,koch_Demonstrating_2020}. 
One strategy to protect quantum states against decoherence is quantum error correction (QEC), which is well suited to mitigating local Markovian noise \cite{shor_Scheme_1995,steane_Error_1996,terhal_Quantum_2015}.
Temporal and spatial noise correlations are however generically hostile to QEC \cite{klesse_Quantum_2005,clader_Impact_2021,clemens_Quantum_2004}, although there are exceptions for weak correlations \cite{terhal_Faulttolerant_2005,aharonov_FaultTolerant_2006}.
Dynamical decoupling (DD), another common noise mitigation strategy, is effective against slow, non-Markovian noise \cite{viola_Dynamical_1998,viola_Dynamical_1999,uhrig_Keeping_2007,cywinski_How_2008,uhrig_Exact_2008,medford_Scaling_2012,zhang_Protected_2014}; however, standard DD does not take advantage of any spatial noise correlations. In systems with many qubits, long range spatial noise correlations have been measured \cite{chwalla_Precision_2007,monz_14Qubit_2011,schindler_Experimental_2011,harper_Efficient_2020,wilen_Correlated_2021}.  There is thus ample motivation for understanding how spatial correlations could be harnessed as a resource for improved noise mitigation.

 \begin{figure}[!t]
     \centering
    \includegraphics[width=0.99\columnwidth]{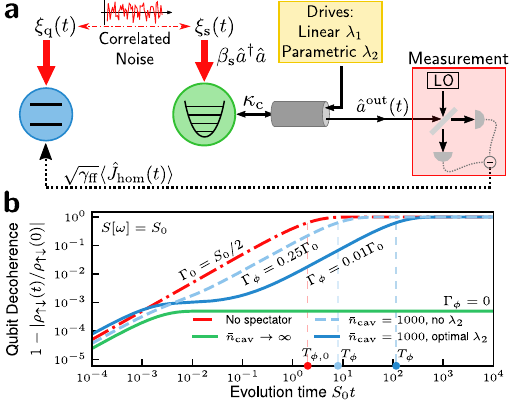}
     \caption{
        \textbf{Spectator photonic mode for mitigating correlated classical noise.} 
        \textbf{(a)} 
        A qubit is dephased by classical noise $\xi_{\rm q}(t)$, which is correlated with the frequency noise  $\xi_{\rm s}(t)$ of a driven photonic spectator mode.
        The spectator's output field is continuously measured, and the resulting measurement record is fedforward to the qubit to mitigate $\xi_{\rm q}(t)$.
        \textbf{(b)}
        Qubit decoherence $1 - |\rho_{\uparrow\downarrow}(t)/\rho_{\uparrow\downarrow}(0)|$ vs. time for perfectly correlated white noise, $S[\omega]=S_0$,
        with and without spectator mode mitigation.  Without the spectator (red dashed curve), the qubit dephases at the rate $\Gamma_0 = S_0/2$.
        With spectator mitigation, the dephasing can be highly suppressed.  Results are shown for $\nbartoti = 1000$ intracavity photons, with and without a parametric drive $\lambda_2$.  For these parameters, an optimized $\lambda_2$ dramatically suppresses dephasing.  
        We also plot the case  $\nbartoti\to\infty$, which completely suppresses the long-time dephasing rate. 
        (In the $\nbartoti\to\infty$ limit, the imprecision noise is zero so squeezing does not improve performance; thus we take $\lambda_2=0$.)
        The dephasing time $T_\phi$ (when the coherence falls by $1/{\rm e}$) is shown for each curve.
        The parameters are $\detr=0.5$, $S_0/\kappa_{\rm c} = 0.001$, and $\as = 1$; the noise strength satisfies Eq.~\eqref{eq:linear-valid} and the squeezing $\lambda_2 \approx 0.74$ satisfies Eq.~\eqref{eq:squeezing-condition}.
        }
     \label{fig:intro}
 \end{figure} 

To this end, there has been a flurry of activity developing spectator qubits (SQ) protocols which explicitly use spatial correlations to fight noise \cite{gupta_Adaptive_2020,gupta_Integration_2020,majumder_Realtime_2020,song_Optimized_2023,singh_Midcircuit_2023}.
The SQ are a dedicated set of qubits in a quantum processor or register which do not interact with the data qubits -- those whose states are to be protected -- but are in close enough physical proximity to be susceptible to the same noise.
By making appropriate measurements of the SQ, one can obtain information about the noise in real time and use this information to apply corrective controls to the data qubits. 
Recent advances in the fabrication and control of many-qubit devices 
\cite{arute_Quantum_2019,koch_Demonstrating_2020} suggests that the use of qubits as spectators could be an attractive approach when spatial noise correlations are present.
Nevertheless, there are some important limitations to the use of SQ.
First, the SQ strategy requires strong spatial noise correlations to work; in fact most existing theory work on SQ has assumed perfect noise correlation.
Second, there are limitations associated with measurement noise: one cannot perfectly estimate the correlated environmental noise from a finite set of measurements of the SQ \cite{majumder_Realtime_2020}.  These unavoidable measurement errors corrupt subsequent correction pulses applied to the target data qubits, ultimately reducing their coherence.
As argued in Ref.~\cite{majumder_Realtime_2020} on information theoretic grounds and taken as an necessary starting assumption in Ref.~\cite{song_Optimized_2023}, to have the SQ scheme be effective despite a finite measurement imprecision, one must make the SQ much more sensitive to the correlated noise than the data qubits.

In this paper we introduce an alternate approach to the spectator philosophy that alleviates many of these problems.
Instead of a qubit, we use a driven mode of a photonic cavity as the spectator quantum system: it detects and mitigates the classical dephasing noise affecting the qubit (see  Fig.~\ref{fig:intro}(a)).
The use of a multi-level photonic mode as the spectator quantum system dramatically suppresses the measurement imprecision problem, for the simple reason that one can now use many photons to estimate the noise. 
As we show, the only way to achieve such low measurement imprecision using SQ would be to measure a large number of SQ, something that is infeasible in many systems.   
Our analysis also reveals another advantage of using a spectator photonic mode: by parametrically driving the spectator mode, the resulting squeezing-induced reduction of the measurement imprecision can exhibit Heisenberg-limited scaling in the number of measurement photons used  \cite{giovannetti_QuantumEnhanced_2004}.

In addition to analyzing a spectator cavity rather than a qubit, there is another crucial difference in the setup we consider.
Whereas standard SQ schemes involve repeated discrete measurements, here we consider an approach based on weak continuous measurements, something that is generally easier and more natural for photonic modes \cite{wiseman_Quantum_2009}. 
The photons in the mode are allowed to leak into a waveguide which is continuously monitored (e.g.~via a homodyne measurement), producing a continuous measurement record that reflects the correlated noise of interest.
Assuming the data qubit is frequency tunable, this measurement record is continuously fedforward to the data qubit, modulating its frequency to correct the noise-induced dephasing.  
This represents another application of continuous feedback control via weak continuous measurements \cite{hofmann_Quantum_1998,korotkov_Selective_2001,gillett_Experimental_2010,vijay_Stabilizing_2012}.
Perhaps even more interesting is that the continuous measurement and feedforward noise mitigation strategy we depict could also be implemented in a {\it completely autonomous} fashion.  In such an approach, no explicit measurements or processing of a classical measurement record are needed; instead, one engineers effective dissipation that mimics the effects of the feedforward process \cite{metelmann_Nonreciprocal_2017}. 

Our analysis has other salient features.  
We go beyond the idealized situation assumed in most previous studies of perfect noise correlation, and explicitly consider the impact of partial noise correlations between the spectator mode and data qubit.  We also consider environmental noise with an arbitrary noise spectral density (as opposed to focusing on one specific form).
Our analysis reveals that the spectator mode approach can still be useful even with partial noise correlations, and for almost any kind of noise spectrum.  In particular, it can even help ameliorate the effects of white noise, despite the finite bandwidth of the feedforward dynamics.  Mitigation of white noise is shown in Fig.~\ref{fig:intro}(b).

The remainder of this paper is organized as follows. In Sec.~\ref{sec:setup} we discuss the detailed setup of the spectator mode, the frequency-tunable qubit, and the measurement and feedforward quantum master equation describing the composite system. In Sec.~\ref{sec:noise-mitigation} we discuss the noise mitigation properties of the spectator mode under both perfect and partial noise correlation. In Sec.~\ref{sec:Heisenberg-limit} we show that the measurement imprecision exhibits Heisenberg-limited scaling in the number of photons used in the measurement, and discuss practical considerations for minimizing measurement imprecision. We discuss applications to superconducting circuits platforms and conclude in Sec.~\ref{sec:conclusion}.


\section{Results}

\subsection{Physical setup}\label{sec:setup}

The spectator mode setup is depicted schematically in Fig.~\ref{fig:intro}(a).
The frequency-tunable qubit is coupled to the classical noise $\xi_{\rm q}(t)$, which modulates its frequency.
The time-dependent qubit Hamiltonian is
\begin{align}
    \hat{H}_{{\rm q},[\xi]}(t) = \frac{1}{2}\left[ \Omega_{\rm q}(\Phi) + \xi_{\rm q}(t) \right]\hat{\sigma}_{\rm z},
    \label{eq:Hq}
\end{align}
where $\hat{\sigma}_{\rm z} = {\lvert\uparrow\rangle\langle\uparrow\rvert} - {\lvert\downarrow\rangle\langle\downarrow\rvert}$ for qubit energy eigenstates ${\lvert\uparrow\rangle},{\lvert\downarrow\rangle}$ and $\Omega_{\rm q}(\Phi)$ is the qubit splitting frequency controlled by some external parameter $\Phi$. 
Here the $[\xi]$ subscript denotes quantities that are functions of noise variables $\xi(t)$. 
We work in a rotating frame about the static qubit splitting frequency set by the operating point $\Phi_0$.
Thus $\Omega_{\rm q}(\Phi_0)\equiv 0$ in Eq.~\eqref{eq:Hq}.

The spectator photonic mode is dispersively coupled to the classical noise $\xi_{\rm s}(t)$, and is continuously driven by a \emph{required} linear drive and an \emph{optional} parametric drive. 
These drives are resonant with the spectator mode frequency $\omega_0$. 
In the rotating frame at $\omega_0$, the time-dependent spectator mode Hamiltonian is given by
\begin{align}
    \hat{H}_{{\rm spec},[\xi]}(t) = \detr \xi_{\rm s}(t)\hat{a}^\dagger\hat{a} + \hat{H}_{\rm drive},
    \label{eq:Hspec}
\end{align}
where $\detr = \partial \omega_{\rm cav}/\partial \xi_{\rm s}$ is a \emph{dimensionless} coupling factor (i.e.~how strongly does the spectator see the noise), and $\hat{H}_{\rm drive}$ is given by Eq.~\eqref{eq:Hdrive}, discussed in detail below.

 \begin{figure}[t]
     \centering
    \includegraphics[width=0.99\columnwidth]{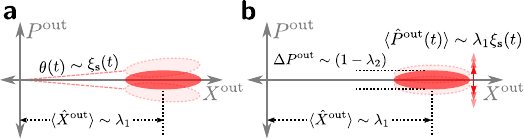}
     \caption{
        \textbf{Phase space representation of spectator output field.} 
        The spectator drives $\hat{H}_{\rm drive}$ (cf. Eq.~\eqref{eq:Hdrive}) create a displaced squeezed state in the cavity that is continuously emitted into the output field $\hat{a}^{\rm out}$ (cf. Eq.~\eqref{eq:aout-relation}). 
        The linear drive $\propto\lambda_1$  displaces the output field along the amplitude quadrature ($\hat{X}$) and the parametric drive $\propto\lambda_2$ squeezes the phase quadrature ($\hat{P}$).
        \textbf{(a)} The dispersive coupling to $\xi_{\rm s}(t)$ in Eq.~\eqref{eq:Hspec} rotates the output field by a small angle $\theta(t)\sim\xi_{\rm s}(t)/\kappa_{\rm c}$ in phase space. 
        To first order in small $\theta$, the rotation causes displacement in the phase quadrature with negligible rotation of the squeezed state.
        \textbf{(b)} After making the linear noise drive approximation (cf. Eqs.~\eqref{eq:Hspecprime} and \eqref{eq:Hs}), the noise is imprinted as fluctuations in the phase quadrature of the output field.}
     \label{fig:aout-blob}
 \end{figure} 

Our noise model is classical stationary Gaussian noise with an arbitrary spectral density $S[\omega]$.
We also consider a generic situation where the qubit noise $\xi_{\rm q}(t)$ and spectator noise $\xi_{\rm s}(t)$ have identical spectral densities (reflecting translational invariance, i.e., the noise is the same at the spectator and data qubit positions):
\begin{align}
    \overline{\xi_{\rm q}(t)\xi_{\rm q}(0)} = \overline{\xi_{\rm s}(t)\xi_{\rm s}(0)} = \int\frac{d\omega}{2\pi} {\rm e}^{-{\rm i}\omega t}S[\omega].
    \label{eq:noise-spec-iden}
\end{align}
The notation $\overline{\bullet}$ indicates the ensemble average over the classical noise variables $\xi(t)$, and $\langle\bullet\rangle$ indicates a quantum expectation value without averaging over $\xi(t)$.
$\xi_{\rm q}(t)$ and $\xi_{\rm s}(t)$ will not in general be perfectly correlated.
Partial correlation between the two noise sources is encoded in the cross-correlation
\begin{align}
    \overline{\xi_{\rm s}(t)\xi_{\rm q}(0)} = \eta \int\frac{d\omega}{2\pi} {\rm e}^{-{\rm i}\omega t}S[\omega],
    \label{eq:etaDefinition}
\end{align}
where $\eta$ parameterizes the degree of correlation, $0\leq\eta\leq1$ (negative correlation $\eta < 0$ requires inverting the signal fedforward to the qubit).
Here we are considering a simple model of partially correlated noise: $\xi_{\rm s}(t)$ is a linear combination of $\xi_{\rm q}(t)$ and another independent noise source with the same spectral density.

We focus on classical noise here because it captures the essential physics of correlated dephasing noise.
One may consider, e.g., a model of quantum dephasing noise which has an asymmetric noise spectral density; however, dephasing noise depends only on the symmetrized spectral density $(S[\omega]+S[-\omega])/2$ and the quantum effects appear only as additional correlated dissipation \cite{seif_Distinguishing_2022}.
Further work is needed to treat the effects of quantum noise.
Similarly, there may be physical situations wherein properties of the noise such as the correlation $\eta$ or noise spectral density $S[\omega]$ could vary in time; however, such scenarios would require a non-stationary noise model, which is beyond the scope of this work.

The linear and parametric drives applied to the spectator mode are described in Eq.~\eqref{eq:Hspec} by the following drive Hamiltonian (in the rotating frame of the mode):
\begin{align}
    \hat{H}_{\rm drive} = \frac{{\rm i}\kappa_{\rm c}}{2}\left[ \lambda_1 \left(\hat{a}^{\dagger}-\hat{a}\right)+\frac{\lambda_2}{2}\left(\hat{a}^{\dagger2}-\hat{a}^{2}\right)\right].
    \label{eq:Hdrive}
\end{align}
We choose to parameterize the drive amplitudes by the rate $\kappa_{\rm c}$, which will be the coupling rate between the mode and an external waveguide (cf. Fig.~\ref{fig:intro}(a)). 
The dimensionless real parameters $\lambda_1,\lambda_2>0$ are the linear drive strength and parametric drive strength, respectively, and $\lambda_2<1$ to avoid parametric instability. 
We assume that there is negligible internal loss in the spectator mode, $\kappa_{\rm i}\ll\kappa_{\rm c}$, thus the mode is overcoupled to the external waveguide such that the total damping rate is $\kappa_{\rm tot} = \kappa_{\rm c}$.

The first drive we apply is \emph{required} to implement the measurement: it is a linear drive $\lambda_1$ in Eq.~\eqref{eq:Hdrive} that displaces the mode from vacuum.
With our phase choice, the drive defines the amplitude quadrature [$\hat{X} \equiv (\hat{a}^\dagger + \hat{a})/\sqrt{2}$] of the mode. 
The displacement causes the frequency fluctuations $\propto\xi_{\rm s}(t)$ to appear as fluctuations in the corresponding phase quadrature [$\hat{P} \equiv {\rm i}(\hat{a}^\dagger - \hat{a})/\sqrt{2}$] (cf. Fig.~\ref{fig:aout-blob}(a)). 
These phase quadrature fluctuations are imprinted on the light that leaks into the waveguide.

The second drive we apply is an \emph{optional} parametric drive $\lambda_2$ in Eq.~\eqref{eq:Hdrive} that, with our phase choice, squeezes the phase quadrature of the mode, as well as the phase quadrature of the mode's output field.
Further, the parametric drive increases the effective damping rate of the intracavity phase quadrature to
\begin{align}
    {\kappa_\phi} \equiv (1+\lambda_2)\kappa_{\rm c}.
    \label{eq:kappa-phi}
\end{align}
Recent work has shown that this kind of in-situ squeezing generation can enhance parameter-estimation measurements, despite the modification of the cavity susceptibility \cite{levitan_Dispersive_2016,peano_Intracavity_2015,korobko_Beating_2017,eddins_HighEfficiency_2019}.  As we show, the same will be true in our setup (cf. Fig.~\ref{fig:aout-blob}(b)), and will allow our spectator scheme to achieve a Heisenberg-limit scaling in the number of measurement photons used.

It is convenient to make the displacement transformation $\hat{a} = \hat{d} + \sqrt{\nbardi}$ in terms of the average photon number associated with the average mode amplitude $\langle\hat{a}\rangle$:
\begin{align}
    \nbardi \equiv
    \left| \langle \hat{a} \rangle \right|^2 = 
    \frac{\lambda_1^2}{(1-\lambda_2)^2}.
    \label{eq:displacement-nbar}
\end{align}
In the displaced frame, the spectator Hamiltonian Eq.~\eqref{eq:Hspec} becomes
\begin{align}
    \label{eq:Hspecprime}
    \hat{H}_{{\rm spec},[\xi]}^{\prime}(t) =&~\detr\xi_{\rm s}(t)\sqrt{\nbardi}(\hat{d}+\hat{d}^\dagger) \\ 
    & - \frac{{\rm i}\kappa_{\rm c}}{4}\lambda_2 (\hat{d}^{2}-\hat{d}^{\dagger2}) 
    + \hat{H}_{{\rm freq},[\xi]}(t), \nonumber \\
    \hat{H}_{{\rm freq},[\xi]}(t) =&~\detr\xi_{\rm s}(t)\hat{d}^\dagger\hat{d}.
    \label{eq:Hfreq}
\end{align}
In this frame, the noise couples to the mode in two ways.  The first is via an effective linear drive $\propto  \sqrt{\nbardi} \xi_{\rm s}(t)$ that displaces the cavity's phase quadrature.  It is this driving that we wish to exploit.  
The second coupling is a quadratic spurious phase noise term $\hat{H}_{{\rm freq},[\xi]}(t)$ that has no $\sqrt{\nbardi}$ enhancement factor.  We wish to work in regimes where this coupling is negligible, as without it, the phase quadrature will have a simple linear dependence on $\xi_{\rm s}(t)$, greatly simplifying our spectator scheme.  

A careful analysis lets us identify regimes where the effects of $\hat{H}_{{\rm freq},[\xi]}(t)$ can be neglected (see Methods Sec.~\ref{app:linearization}).  
In the case where no squeezing is employed  ($\lambda_2=0$), we require: 
\begin{align}
    \detr^2\int &\frac{d\omega}{2\pi} \frac{S[\omega]}{\omega^2} \sin^2(\omega/2 \kappa_{\rm c}) \ll 1. \label{eq:linear-valid}
\end{align}
Heuristically, this condition ensures that the phase diffusion induced by the spurious phase noise term is negligible during the relevant correlation time of the cavity mode.  
For example, in the case of white noise $S[\omega]=S_0$, Eq.~\eqref{eq:linear-valid} reduces to the constraint $\detr^2 S_0 \ll \kappa_{\rm c}$.

It may seem counterintuitive that the spectator should be weakly sensitive to the correlated noise -- in the sense that $\detr$ should be small in some sense, as one might expect that the spectator mode should be maximally sensitive to the noise.
It is important to note that in Eq.~\eqref{eq:Hspec}, the ``good'' (linear) coupling to the noise is controlled by $\detr\nbardi$ whereas the spurious nonlinear coupling strength is controlled by only $\detr$. 
Eq.~\eqref{eq:linear-valid} (or Eq.~\eqref{eq:squeezing-condition} below) gives the regime for which the spurious coupling can be neglected.
Since these constraints do not involve $\nbardi$, it remains a control knob for the sensitivity to the good linearly coupled noise.

The spurious phase noise term also constrains the use of squeezing in our scheme, i.e., a non-zero $\lambda_2$.   
Heuristically, this unwanted dynamics induces a rotation in phase space that mixes the enhanced amplitude quadrature quantum noise into the squeezed phase quadrature \cite{zhang_Quantum_2003,aoki_Squeezing_2006,oelker_Ultralow_2016}.
To have this extra noise not overwhelm the desired noise squeezing, the following condition must also be satisfied:
\begin{align}
    (1-\lambda_2)^4 \gtrsim 16 \detr^2\int &\frac{d\omega}{2\pi} \frac{S[\omega]}{\omega^2} \sin^2(\omega/2 \kappa_{\rm c}). \label{eq:squeezing-condition}
\end{align}
See Methods Sec.~\ref{app:linearization} for a detailed discussion.  We stress that even when Eq.~\eqref{eq:linear-valid} is satisfied, this condition determines the maximum value of $\lambda_2$ (and hence squeezing) that can be usefully employed to enhance noise mitigation.

In what follows, we assume both that $\nbardi \gg 1$ and that the noise is weak enough that both Eqs.~\eqref{eq:linear-valid} and \eqref{eq:squeezing-condition} hold.  We can thus safely approximate the spectator mode Hamiltonian as
\begin{align}
    \hat{H}_{{\rm s},[\xi]}(t) &\simeq \detr\xi_{\rm s}(t)\sqrt{\nbardi}(\hat{d}+\hat{d}^\dagger) - \frac{{\rm i}\kappa_{\rm c}}{4}\lambda_2 (\hat{d}^{2}-\hat{d}^{\dagger2}),
    \label{eq:Hs}
\end{align}
i.e.~the spectator is only linearly driven by the noise.

In an experiment, one may ensure that Eq.~\eqref{eq:linear-valid} (or Eq.~\eqref{eq:squeezing-condition} if using squeezing) holds by the combination of engineering a sufficiently weak dispersive coupling strength of the noise to the spectator mode (reducing $\detr$) and increasing the coupling of the spectator mode to the external waveguide, $\kappa_{\rm c}$.
For example, suppose the spectator mode were a superconducting cavity that detects spurious magnetic fields using a SQUID loop.
Through a combination of setting the geometric area of the SQUID loop during fabrication and biasing the SQUID loop with a DC flux, one can control the sensitivity of the spectator mode to the noise, $\detr = \partial \omega_{\rm cav}/\partial \Phi_{\rm ext}$ in-situ.

The final elements of our setup are the spectator measurement and feedforward operations.  While we start by analyzing a protocol that involves an explicit continuous measurement, we will end with an effective description that could be implemented in a fully autonomous manner (i.e.~without needing any explicit measurement or processing of a classical measurement record).  

Via Eq.~\eqref{eq:Hs}, the noise $\xi_{\rm s}(t)$ modulates the phase of the light leaving the cavity (see Fig.~\ref{fig:aout-blob}(b)).
This information is captured in the output field phase quadrature:
\begin{align}
    \hat{P}^{\rm out}_{[\xi]}(t) &= \frac{\rm i}{\sqrt{2}}\left(\hat{a}^{{\rm out}\dagger}_{[\xi]}(t) - \hat{a}^{{\rm out}}_{[\xi]}(t)  \right), \\ 
    \hat{a}^{\rm out}_{[\xi]}(t) &= \hat{a}^{\rm in}(t) + \sqrt{\kappa_{\rm c}}\hat{a}_{[\xi]}(t).
    \label{eq:aout-relation}
\end{align}
This output field phase quadrature can be continuously measured using a standard homodyne measurement (see e.g., \cite{scully_Quantum_1997,gerry_Introductory_2004}), producing the continuous homodyne current $\hat{J}_{{\rm hom},[\xi]}(t) = \hat{P}^{\rm out}_{[\xi]}(t)$. 
To implement our noise mitigation, we will use this homodyne current (which has units of $\sqrt{\rm rate}$) for feedforward control of the qubit, namely to directly modulate the qubit frequency. 
This amounts to using the measurement record $\sim\langle\hat{J}_{{\rm hom},[\xi]}\rangle$ to modulate the external parameter $\Phi$ that controls the qubit frequency in Eq.~\eqref{eq:Hq}.  Ignoring delays, we obtain a modified qubit Hamiltonian
\begin{align}
    \hat{H}_{{\rm q},[\xi]}^{\prime}(t) = \frac{1}{2}\left( \xi_{\rm q}(t) + \sqrt{\gff}\hat{J}_{{\rm hom},[\xi]}(t) \right) \hat{\sigma}_{\rm z},
    \label{eq:homodyne-currrent-H}
\end{align}
where the rate $\gff$ parameterizes the strength of the feedforward. 

If the spectator were a perfect, noiseless classical system with an instantaneous response, we could tune the feedforward strength to achieve $\sqrt{\gff}J_{{\rm hom},[\xi]}(t) \rightarrow -\xi_{\rm q}(t)$.  The feedforward would thus cancel the noise experienced by the qubit in Eq.~\eqref{eq:homodyne-currrent-H} completely.  For a realistic quantum spectator mode, this perfect cancellation is degraded by two basic effects.  First, the spectator mode always has a non-instantaneous response which limits its sensitivity to high-frequency noise.  Second, the homodyne current is operator-valued and thus has quantum fluctuations which add unwanted quantum noise to the measurement record.
The added noise is effectively a random error at each time step, thus it is the continuous analog of imprecision errors in a set of discrete measurements.
The fully quantum description we develop below allows us treat both the limitations due to quantum noise and the delayed spectator response. 

Using standard continuous measurement theory \cite{wiseman_Quantum_2009}, the unconditional dynamics of our feedforward setup (i.e.~averaged over measurement outcomes) can be described using a Lindblad master equation that governs the combined spectator mode and qubit dynamics. 
As the mode is included in our description, we retain all effects associated with its non-instantaneous response.  
The resulting master equation is still a function of the specific classical noise realizations $\xi_j(t)$ (which need not be white noise), and takes the form:
\begin{align}
    \frac{d}{dt}\hat{\rho}_{[\xi]}(t) = &-{\rm i}[\hat{H}_{{\rm q},[\xi]}(t) + \hat{H}_{{\rm s},[\xi]}(t) + \hat{H}_{\rm int},\hat{\rho}]+\mathcal{D}[\hat{L}]\hat{\rho},
    \label{eq:qme}
\end{align}
where $\mathcal{D}[\hat{X}]\hat{\rho} = \hat{X}\hat{\rho}\hat{X}^\dagger - \{\hat{X}^\dagger\hat{X},\hat{\rho}\}/2$ is the standard Lindblad dissipator, 
$\hat{H}_{{\rm q},[\xi]}(t)$ is given by Eq.~\eqref{eq:Hq}, and $\hat{H}_{{\rm s},[\xi]}(t)$ is given by Eq.~\eqref{eq:Hs}. 
The measurement and feedforward induces a Hamiltonian interaction and collective dissipation given by
\begin{align}
    \hat{H}_{\rm int} &= \frac{1}{2{\rm i}}\sqrt{\gff\kappa_{\rm c}}\left(\hat{d}-\hat{d}^\dagger\right)\hat{\sigma}_{\rm z},
    \label{eq:Hint} \\
    \hat{L} &= \sqrt{\kappa_{\rm c}}\hat{d} + \sqrt{\gff}\hat{\sigma}_{\rm z}.
    \label{eq:L-diss}
\end{align}

\begin{figure}[t]
     \centering
    \includegraphics[width=0.999\columnwidth]{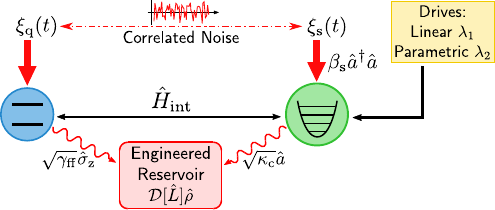}
     \caption{
     \textbf{Autonomous spectator mode noise mitigation.}
     The autonomous implementation of the spectator mode requires a direct Hamiltonian interaction $\hat{H}_{\rm int}$ (cf. Eq.~\eqref{eq:Hint}) between the spectator and the qubit as well as an engineered reservoir that mediates the collective dissipation $\mathcal{D}[\hat{L}]\hat{\rho}$ (cf. Eq.~\eqref{eq:L-diss}).
     These elements replace the measurement and feedforward apparatus shown in Fig.~\ref{fig:intro}(a), and could be engineered in cavity QED setups using auxiliary cavity modes, parametric couplings and external drives. See Methods Sec.~\ref{app:autonomous} for more details.
     }
     \label{fig:autonomous}
 \end{figure}

We pause to make an important comment on Eq.~\eqref{eq:qme}: while this dissipative dynamics can be generated (as discussed) via continuous measurement and feedforward, it could also be directly generated without any measurement, using instead the tools of reservoir engineering.  This autonomous approach would replace the measurement and feedforward parts of our setup by an engineered dissipative bath that couples to both the qubit and spectator \cite{metelmann_Nonreciprocal_2017}, in such a way to realize Eq.~\eqref{eq:qme} (see Fig.~\ref{fig:autonomous}).  We provide more details in Methods Sec.~\ref{app:autonomous} showing how such a reservoir could be constructed using standard tools in cavity QED. This autonomous realization is a potentially powerful approach to spectator-based noise mitigation, as it does not require high fidelity measurements nor any interface with the classical world beyond the pump tones needed to implement the spectator mode and reservoir engineering.

The qubit evolution governed by Eq.~\eqref{eq:qme} conserves $\hat{\sigma}_{\rm z}$, and thus it has pure dephasing dynamics.  
Suppose the qubit is prepared in the superposition state $\lvert\psi_{\rm q}(0)\rangle = (\lvert\uparrow\rangle + \lvert\downarrow\rangle)$.
The noise can only change the superposition phase as $\lvert\psi_{\rm q}(t)\rangle = (\lvert\uparrow\rangle + \exp{[-{\rm i}\int_0^t ds\xi_{\rm q}(s)]} \lvert\downarrow\rangle)$.
We are thus interested in studying the decay of the qubit coherence due to this random phase accumulation. The qubit coherence is given by the noise-averaged off-diagonal matrix element of the qubit density matrix
\begin{align}
    \rho_{\uparrow\downarrow}(t) =
    \overline{\langle\uparrow\rvert\hat{\rho}_{\rm q}(t) \lvert\downarrow\rangle},
    \label{eq:rho-up-down-def}
\end{align}
where $\lvert\uparrow\rangle,\lvert\downarrow\rangle$ are the eigenstates of $\hat{\sigma}_{\rm z}$ and $\hat{\rho}_{\rm q}(t)$ is the qubit density matrix. The loss of qubit coherence is characterized by a decay in the magnitude of $\rho_{\uparrow\downarrow}(t)$ with time.

We will parameterize the qubit coherence by the decoherence function $\chi(t)$, defined via
\begin{align}
    \rho_{\uparrow\downarrow}(t) = \rho_{\uparrow\downarrow}(0){\rm e}^{-\chi(t)}.
\end{align}
In the simple case of Markovian dephasing (i.e.~due to white noise), we have $\chi(t) = \Gamma_\phi t$, where $\Gamma_\phi$ is the linear dephasing rate.
\emph{Without the spectator system}, the bare qubit decoherence function $\chi_0(t)$ is given by
\begin{align}
    \chi(t) \to \chi_0(t)
    = \frac{1}{2}\int\frac{d\omega}{2\pi}\frac{S[\omega]}{\omega^2} |Y_{\rm fid}(\omega,t)|^2
\end{align}
where we have used the Gaussian nature of the environmental noise (spectrum $S[\omega]$), and  the free induction decay (FID) filter function $Y_{\rm fid}(\omega,t)$ is
\begin{align}
    Y_{\rm fid}(\omega,t) = {\rm e}^{-{\rm i}\omega t} - 1.
    \label{eq:fid-filter-func}
\end{align}

To find the decoherence function when the spectator mode is introduced, we compute the stochastic qubit coherence $\rho_{\uparrow\downarrow,[\xi]}(t)$, then average over noise realizations.
The qubit reduced density matrix is given by $\hat{\rho}_{{\rm q},[\xi]}(t) = {\rm tr}_{\rm s}\{\hat{\rho}_{[\xi]}(t)\}$, where the partial trace is over the spectator mode, and $\hat{\rho}_{[\xi]}(t)$ evolves under Eq.~\eqref{eq:qme}.
We find that
\begin{align}
    \rho_{\uparrow\downarrow,[\xi]}(t) = \rho_{\uparrow\downarrow}(0){\rm e}^{-{\rm i}\phi_{[\xi]}(t)}{\rm e}^{-\LamImp(t)}
    \label{eq:rho-up-down}
\end{align}
where $\phi_{[\xi]}(t)$ is the total stochastic phase accumulation, and $\LamImp(t) \geq 0$ is the ``measurement imprecision noise dephasing'' i.e., the qubit decoherence due to quantum noise associated with the measurement-plus-feedforward dynamics.  As we will show, this decoherence is a direct consequence of the imprecision noise associated with the spectator ``measuring'' the environmental noise.  

First, consider the accumulated phase $\phi_{[\xi]}(t)$.  We find:
\begin{align}
	\phi_{[\xi]}&(t)= \label{eq:phi}\\
	&\int_{0}^{t}dt^{\prime}\left[\xi_{\rm q}(t^{\prime}) -  \as\frac{\kappa_\phi}{2}\int_{-\infty}^{t^{\prime}}ds {\rm e}^{-{\kappa_\phi} (t^{\prime}-s)/2}\xi_{\rm s}(s)\right],\nonumber
\end{align}
where the $\xi_{\rm q}(t^{\prime})$ term is the direct noise on the qubit and the $\xi_{\rm s}(s)$ term is the filtered, fedforward noise driving the spectator. This expression and $\LamImp(t)$ (given below) are derived in Methods Sec.~\ref{app:decoherence-func}. The parameter $\as$ is the dimensionless spectator transduction factor
\begin{align}
    \as &\equiv \frac{8\detr}{\kappa_{\phi}}\sqrt{\nbardi\kappa_{\rm c}\gff}\label{eq:alpha-s}.
\end{align}
It controls how the environmental noise sensed by the spectator ultimately drives the qubit.
The transduction strength is set by the spectator detection factor $\detr$ (cf. Eq.~\eqref{eq:Hspec}), spectator mode displacement $\sqrt{\nbardi}$ (cf. Eq.~\eqref{eq:displacement-nbar}), the feedforward rate $\sqrt{\kappa_{\rm c} \gff}$ (cf. Eqs.~\eqref{eq:Hdrive} and \eqref{eq:homodyne-currrent-H}), and the inverse of the phase quadrature damping rate $1/\kappa_\phi$ (cf. Eq.~\eqref{eq:kappa-phi}) -- strong damping requires a stronger transduction strength.
We view $\as$ as the relevant control parameter and henceforth write $\gff$ in terms of $\as$.

One can easily confirm that this expression for the accumulated phase $\phi_{[\xi]}(t)$ has a  completely classical form, i.e.~what we would expect if the spectator were a classical (noiseless) system with a delayed,  exponential response.   
The quantum nature of the spectator mode only appears in the measurement imprecision noise dephasing $\LamImp(t)$, which stems from (squeezed) vacuum noise in the spectator phase quadrature.  We find
\begin{align}
    &\LamImp(t) = \label{eq:lambda-add} \\ 
    &~\frac{\as^2}{32\detr^2 \nbardi}  \frac{1}{1+\lambda_2} \left[(1-\lambda_{2})^{2}\kappa_{\phi}t+8\lambda_{2}\left(1-{\rm e}^{-\kappa_{\phi}t/2}\right)\right] . \nonumber
\end{align}
Without squeezing $\lambda_2=0$, the added quantum dephasing noise is white so $\LamImp(t)$ is linear in time; this is no longer true when $\lambda_2\neq0$. One already sees the advantage of squeezing by the dramatic reduction in the long-time added-noise dephasing:
\begin{align}
    \LamImp(t) = 
    \begin{cases}
        2\gff t & t\ll1/\kappa_\phi \\ 
        \left(\frac{1-\lambda_2}{1+\lambda_2}\right)^2 (2\gff t) + {\rm const} & t\gg1/\kappa_\phi
    \end{cases}.
    \label{eq:LambdaAddLimits}
\end{align}
At short times, we have exponential dephasing set by the feedforward rate, whereas at long times, squeezing can strongly suppress this exponential dephasing.  

Returning to the decoherence function $\chi(t)$, we finally take the ensemble average of Eq.~\eqref{eq:rho-up-down} over realizations of the environmental noise:
\begin{align}
    \rho_{\uparrow\downarrow}(t) =  \rho_{\uparrow\downarrow}(0)\overline{{\rm e}^{{\rm i}\phi_{[\xi]}(t)}}{\rm e}^{-\LamImp(t)}. 
    \label{eq:rho-up-down-avg}
\end{align}
As the accumulated phase is linear in the Gaussian noise $\xi(t)$, the average can be done exactly.  The final form of the decoherence function can thus be written in terms of the spectrum $S[\omega]$ of our environmental noise:
\begin{align}
    \chi(t) &= \LamImp(t)
    \label{eq:dephasing-func} \\
    &+ \frac{1}{2} \int\frac{d\omega}{2\pi} \frac{S[\omega]}{\omega^2}\left[ \lvert Y(\omega, t, \as) \rvert^2 + \lvert \tilde{Y}(\omega, t, \as) \rvert^2\right]. \nonumber
\end{align}
We have introduced the filter functions  $Y(\omega,t,\as)$ and $\tilde{Y}(\omega,t,\as)$, which correspond to the contributions of the correlated and uncorrelated parts of the noise.  
The correlated-noise filter function is
\begin{align}
    Y(\omega, t, \as) &= \left(1 - \eta\as\frac{{\kappa_\phi}/2}{ {\rm i}\omega+{\kappa_\phi}/2}\right) Y_{\rm fid}(\omega,t),
    \label{eq:filter-function-corr}
\end{align}
where $Y_{\rm fid}(\omega,t)$ is given by Eq.~\eqref{eq:fid-filter-func}, and $\eta$ sets the level of noise correlation (c.f.~Eq.~\eqref{eq:etaDefinition}). 
The two terms here correspond to the two ways the environmental noise reaches the qubit: directly (first term) and through the feedforward process (second term).  The spectator scheme relies on getting these terms to cancel as best as possible.  

In contrast, the uncorrelated-noise filter function is
\begin{align}
    \tilde{Y}(\omega,t,\as)&=\sqrt{1-\eta^{2}}\as\frac{{\kappa_\phi}/2}{{\rm i}\omega+{\kappa_\phi}/2}Y_{\rm fid}(\omega,t).
    \label{eq:filter-function-uncorr}
\end{align}
This contribution to the qubit dephasing increases monotonically with $\as$ as expected (i.e.~this is just like an independent new source of noise driving the qubit).


\subsection{Noise mitigation}\label{sec:noise-mitigation}

We start by considering the limit of perfect noise correlation between spectator and data qubit ($\eta=1$); this lets us understand the basic features of our scheme.  
We then show how the noise mitigation is affected by partial noise correlation $\eta<1$.
We stress our treatment includes the delay associated with the spectator mode response; 
in App.~\ref{app:feedforward-delay} and Fig.~\ref{fig:feedforward-delay}, we discuss the additional impact of delay associated with applying the feedforward control on the data qubit.

Our spectator-based noise mitigation ultimately relies on optimizing the cancellation of the two terms in Eq.~\eqref{eq:filter-function-corr} to minimize the dephasing factor in Eq.~\eqref{eq:dephasing-func}.  There is only a single effective parameter to optimize: the transduction factor $\as$, c.f.~Eq.~\eqref{eq:alpha-s}.  One might guess that the optimal value of $\as$ will depend on the form of the noise spectrum $S[\omega]$ and on the evolution time $t$.  This is not the case.  As $\as$ is constrained to be real, a simple calculation shows that there is a single ideal value of $\as$ which minimizes $|Y(\omega,t,\as)|^2$ at all frequencies for all times:
\begin{align}
    \as^{\rm ideal} = 1.
\end{align}
For the perfect correlation case we consider, this in turn minimizes the second term in the dephasing factor $\chi(t)$ in Eq.~\eqref{eq:dephasing-func}, regardless of the time $t$ or form of noise spectrum.  Note this optimal value matches what we would naively expect if there were no delay in the spectator mode response.  

The filter function at ideal transduction strength ($\as=1$) takes the simple form:
\begin{align}
   Y(\omega, t, \as=1) &= \frac{{\rm i} \omega/2}
   { {\rm i}\omega+{\kappa_\phi}/2}  Y_{\rm fid}(\omega,t).
    \label{eq:IdealFilterFunc} 
\end{align}
As expected, the filter function is strongly suppressed over the no-feedforward case at low frequencies $\omega \lesssim \kappa_\phi$, with a perfect suppression at zero frequency.  This latter property implies that the environmental noise will not give any long-time exponential dephasing of the qubit.

A key feature of the spectator mode approach is its ability to mitigate the effects of white noise.  This is directly related to a property of its filter function: its total spectral weight, given by
\begin{align}
    \int \frac{d\omega}{2\pi} &|Y(\omega,t,\as)|^2  = \\
    &\left( 1 - \as \right)^{2} t + \frac{2}{\kappa_\phi} \as (2 - \as)( 1 - {\rm e}^{-\kappa_\phi t/2}),\nonumber
\end{align}
is not fixed. 
In particular, this weight can be made arbitrarily small for the optimal tuning $\as = 1$ by letting the spectator detection bandwidth $\kappa_\phi$ be arbitrarily large.  This is in stark contrast to typical dynamical decoupling spectral functions that conserve spectral weight:
\begin{align*}
    \int \frac{d\omega}{2\pi} \frac{1}{\omega^2} |Y_{\rm dd}(\omega,t)|^2 = \int \frac{d\omega}{2\pi} \frac{1}{\omega^2} |Y_{\rm fid}(\omega,t)|^2.
\end{align*}
The conservation of spectral weight in dynamical decoupling means that one cannot mitigate white noise: the suppression of contributions at one frequency to the dephasing factor necessarily imply increased contributions at other frequencies \cite{biercuk_Dynamical_2011,suter_Colloquium_2016}.  The spectator based approach (which cancels noise induced dephasing via a completely different mechanism) does not have a similar constraint.

 \begin{figure}[t]
     \centering
    \includegraphics[width=0.99\columnwidth]{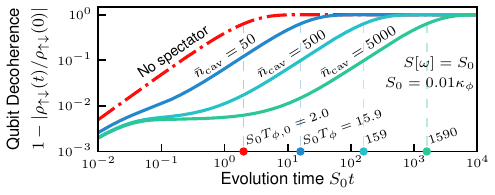}
     \caption{
        \textbf{Spectator cavity performance without squeezing.}
        Qubit decoherence $1-|\rho_{\uparrow\downarrow}(t)/\rho_{\uparrow\downarrow}(0)|$ vs. time for perfectly correlated white noise, $S[\omega]=S_0$, with and without spectator mode mitigation. The characteristic dephasing time $T_\phi$, defined as the time at which the qubit coherence falls by a factor $1/{\rm e}$ (i.e., when the decoherence function $\chi(T_\phi) = 1$, cf. Eq.~\eqref{eq:dephasing-func}), is shown for each curve. Without the spectator mode (red dashed curve), the qubit linearly dephases, losing $1/e$ of its coherence at time $T_{\phi,0} = 2/S_0$. Results are shown with only a one-photon drive applied to the spectator mode. The number of intracavity photons $\nbartoti = \nbardi$ (cf. Eq.~\eqref{eq:displacement-nbar}) is indicated above each curve. The spectator mode can dramatically increase the coherence time of the qubit even without the two-photon drive applied. The parameters are $\detr=1$, $S_0/\kappa_\phi=0.01$, $\lambda_2 = 0$ and $\as = 1$; the noise strength satisfies Eq.~\eqref{eq:linear-valid}.
        }
     \label{fig:squeezing-free-performance}
 \end{figure} 

We now analyze the ultimate performance of the spectator mode approach in mitigating qubit dephasing.  We first consider the asymptotic long-time limit $t \rightarrow \infty$.  As $Y_{\rm fid}(\omega,t)$ approaches a $\delta$ function in this limit, for non-singular spectral densities, the long-time decoherence for any $\as$ (but assuming $\eta = 1$) is:
\begin{align}
    \chi(t\to \infty) &= \frac{1}{2}(\as - 1)^2 S[0]t + \LamImp(t) + \chi_{\textrm{init}}(\infty)  \label{eq:long-time-dephasing}
\end{align}
where $\chi_{\textrm{init}}(\infty)$ is a $t$-independent constant (arising from the finite spectator bandwidth) that we analyze more below.  For an ideal choice of transduction factor $\as=1$, the linear-in-$t$ phase diffusion from the environmental noise is completely cancelled.  In its place, we have the imprecision noise dephasing $\LamImp$ (c.f.~Eq.~\eqref{eq:lambda-add}). Crucially, while $\LamImp$ also grows linearly with $t$ in the long-time limit, the corresponding rate can be made arbitrarily small by using many photons for the measurement.  Defining the long time dephasing rate $\Gamma_\phi = \lim_{t \rightarrow \infty} \chi(t)/t$, and considering the simple case where there is no parametric drive on the spectator cavity (i.e.~$\lambda_2 = 0$) and $\as = 1$, we have both with and without feedforward:  
\begin{equation}
    \left( \Gamma_\phi = \frac{1}{2} S[0] \right) \rightarrow
    \left( \Gamma_\phi = \frac{1}{32 \detr^2 \bar{n}_1} \kappa_{\rm c} \right)
    \label{eq:DephasingRateModification}
\end{equation}
We see that the spectator mode approach (in this ideal limit) suppresses the long time dephasing rate by a factor scaling as $\kappa_{\rm c} /( \bar{n}_1 S[0] )$.  We stress that this result is valid for any noise spectrum that is non-singular at $\omega = 0$, including white noise.

The suppression of the long-time qubit dephasing rate is shown in Fig.~\ref{fig:intro}(b) and in Fig.~\ref{fig:squeezing-free-performance}. In both plots, the qubit is subject to perfectly correlated white noise, $S[\omega]=S_0$. We see in Fig.~\ref{fig:intro}(b) that the use of squeezing, $\lambda_2\neq 0$, dramatically suppresses the long-time dephasing rate relative to the modest suppression achieved using no squeezing (cf. the solid and dashed blue curves of Fig.~\ref{fig:intro}(b)). 
In Fig.~\ref{fig:squeezing-free-performance} we demonstrate that significant suppression of the long-time dephasing rate is achievable without the use of squeezing.
The requirement of sufficiently weak noise strength, Eq.~\eqref{eq:linear-valid}, is much less restrictive when no squeezing is applied. For Fig.~\ref{fig:intro}(b), we must have $S_0/\kappa_\phi \lesssim 0.001$ whereas for Fig.~\ref{fig:squeezing-free-performance}, we require only $S_0/\kappa_\phi \lesssim 0.01$.

In many cases, one is interested in understanding the qubit coherence at all times, not just in the long time limit.  For finite evolution times, the qubit coherence will be sensitive to the environmental noise (and filter function) over a finite bandwidth. Even for an optimal transduction strength $\as = 1$, the relevant filter function is only suppressed (compared to $\as = 0$) for frequencies $\omega \lesssim \kappa_\phi$, c.f.~Eq.~\eqref{eq:IdealFilterFunc}.  Heuristically, this means that even though there is no environmental-noise induced long-time dephasing in this case, there will be some dephasing over a time interval $0 \leq t \lesssim 1 / \kappa_\phi$.  This initial dephasing gives rise to the constant term $\chi_{\rm init}$ in Eq.~\eqref{eq:long-time-dephasing}.

For $\as=1$, the extra contribution to the dephasing factor from finite-frequency environmental noise is given by:
\begin{align}
     \chi_{\textrm{init}}(t)=\frac{1}{2}\int\frac{d\omega}{2\pi} \frac{S[\omega]}{\omega^{2}+\kappa_\phi^{2}/4}|Y_{\rm fid}(\omega,t)|^2.
    \label{eq:chi-init}
 \end{align}
As an example, for white noise $S[\omega]=S_0$, this extra dephasing initially grows linearly in time as $\chi_{\rm init}(t) = S_0 t/2$ for $t\ll 1/\kappa_\phi$, but then it saturates to a constant $\chi_{\rm init}\to S_0/\kappa_\phi$ when $t\sim1/\kappa_\phi$.

\begin{figure}[t]
     \centering
    \includegraphics[width=0.99\columnwidth]{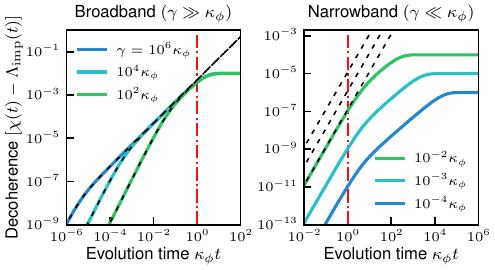}
     \caption{
        \textbf{Finite detection bandwidth effects for broadband and narrowband noise.} The qubit decoherence function (less $\LamImp(t)$) versus time for Lorentzian noise (bandwidth $\gamma$,  zero-frequency strength $S[0] = 0.01\kappa_\phi$). 
        In both plots the black dashed curves are the bare qubit decoherence function (in the right plot, relative vertical position corresponds to same-position labeled curve), and the vertical dot-dashed line indicates $t = 1/\kappa_\phi$. 
        \textbf{Left plot:} broadband noise.  For short times $t<1/\kappa_\phi$ the spectator cannot mitigate the noise so the qubit dephases as it would without the spectator, while for longer times there is strongly suppressed dephasing.
        \textbf{Right plot:} Narrowband noise.  For short times $t<1/\kappa_\phi$ the spectator greatly suppresses the initial quadratic-in-$t$ dephasing. 
        For times $1/\kappa_\phi < t < 1/\gamma$ the spectator reduces dephasing to linear in time with suppressed rate $(\gamma/\kappa_\phi)^2 S_0/2 \ll S_0/2$.
        Dephasing finally saturates after $t>1/\gamma$.}
     \label{fig:short-time-decoherence}
 \end{figure}

The short-time dephasing that arises for $\as = 1$ and a finite spectator bandwidth $\kappa_\phi$ are are illustrated in Fig.~\ref{fig:short-time-decoherence}, for the case of a Lorentzian environmental noise spectrum,  $S[\omega]=S_0(\gamma^2/4)/(\omega^2+\gamma^2/4)$.
As expected, broadband noise ($\gamma\gg\kappa_\phi$) cannot be mitigated over short timescales, hence the spectator-mitigated decoherence closely tracks the bare qubit decoherence until $t\sim1/\kappa_\phi$ (see right plot of Fig.~\ref{fig:short-time-decoherence}).
Around $t\sim1/\kappa_\phi$ the decoherence quickly saturates to a finite value.
The initially quadratic dephasing becomes linear when $t>1/\gamma$.
In contrast, narrowband noise ($\gamma\ll\kappa_\phi$) is much more effectively mitigated by the spectator at all times, as demonstrated by the prefactor suppression of the initial quadratic dephasing (see  right plot of Fig.~\ref{fig:short-time-decoherence}).
At times $t\sim1/\kappa_\phi$ the initially quadratic dephasing becomes linear with a suppressed rate $(\gamma/\kappa_\phi)^2 \Gamma_{0}$.
After a time $t\sim1/\gamma$, $\chi_{\textrm{init}}(t)$ saturates to a finite value.

Our discussion so far has focused on the case of perfect noise correlation between the spectator mode and data qubit, $\eta = 1$.  
When there is only partial noise correlation between the spectator and qubit ($\eta<1$), we must retain both filter functions Eqs.~\eqref{eq:filter-function-corr} and \eqref{eq:filter-function-uncorr} in the expression Eq.~\eqref{eq:dephasing-func} for $\chi(t)$.
The uncorrelated noise prevents the spectator mode from perfectly canceling the long-time dephasing rate, and creates a penalty for increasing the transduction factor $\as$ from $0$.

For a general correlation $\eta$,  the qubit decoherence function in the long-time limit is
\begin{align}
    \chi(t) &= \frac{1}{2}(\as^2 -2\eta\as + 1) S[0]t + \LamImp(t) + {\rm const}.
\end{align}
For imperfect correlation, it is no longer possible to cancel the dominant $S[0] t$ term in this expression.  The best we can do it minimize its prefactor, by tuning the transduction factor to the optimal value
\begin{align}
    \as^{\rm ideal} = \eta,
\end{align}
resulting in
\begin{align}
    \chi(t) &\rightarrow \frac{1}{2}(1-\eta^2 ) S[0]t + \LamImp(t) + {\rm const}. \label{eq:long-time-decoherence-partial-correlation}
\end{align}
Hence, the environmental-noise contribution to the long time dephasing rate can be suppressed by a factor $(1- \eta^2)$, showing that the spectator approach can still be useful even with partial noise correlations.  

Of course, in asking whether spectator based feedforward is beneficial in suppressing long-time dephasing, one also needs to consider the  measurement imprecision noise dephasing $\LamImp$.  The excess dephasing from imprecision noise can always (in principle) be mitigated by using a sufficiently large number of photons $\nbardi$, c.f. Eq.~\eqref{eq:DephasingRateModification}.

The efficacy of the spectator mode at mitigating partially correlated noise is not only a long-time limit result.
One can show that the ideal transduction factor at any time $t$ is $\as^{\rm ideal}=\eta$, and that for this choice of transduction strength the qubit decoherence function is
\begin{align}
    \chi(t) &= \LamImp(t) + \chi_{\rm init}(t) \\
    &+ \frac{1}{2}(1-\eta^{2})\int\frac{d\omega}{2\pi}\frac{S[\omega]}{\omega^{2}}\frac{\kappa_{\phi}^{2}/4}{\omega^{2}+\kappa_{\phi}^{2}/4}|Y_{\mathrm{fid}}(\omega,t)|^{2} \nonumber
\end{align}
where the final term encodes all dephasing due to the partial noise correlation.
Taking white noise $S[\omega]=S_0$ as an example, the qubit decoherence function is found to be
\begin{align}
    \chi(t) &= \LamImp(t) +\eta^{2}\chi_{\rm init}(t) + \frac{1}{2}(1-\eta^{2})S_{0}t,
\end{align}
from which we recover the perfect noise correlation result for $\eta=1$.


\subsection{Measurement imprecision noise dephasing: approaching Heisenberg-limited scaling}
\label{sec:Heisenberg-limit}

As discussed, the key quantum aspect of the spectator mode scheme is the measurement imprecision noise contribution to the qubit's dephasing, $\LamImp(t)$ (c.f.~Eq.~\eqref{eq:lambda-add}).  This term is ultimately due to the (possibly squeezed)
vacuum noise of the spectator mode.  Minimizing this dephasing is a key aspect to the spectator mode strategy. We imagine a situation where the transduction factor $\as$ (c.f.~Eq.~\eqref{eq:alpha-s}) can be fixed to the optimal value of $1$, while the drive amplitudes $\lambda_1$ and $\lambda_2$ can still vary; this can be accomplished by appropriately tuning the feedforward strength $\gff$.
The measurement-imprecision noise dephasing can be made arbitrarily small by using an arbitrarily large number of photons, i.e., let $\lambda_1\to\infty$ and $\lambda_2 \to 1 $.  We wish to understand this more quantitatively, and in a manner more relevant to experiment. Given that we are able to use some fixed total number of photons for the measurement, how small can we make $\LamImp(t)$?  Further, how does this optimized added dephasing scale with photon number?

The above questions are directly connected to quantum limits on parameter estimation \cite{giovannetti_QuantumEnhanced_2004}.
As discussed, $\LamImp(t)$ can be viewed as the variance of the measurement imprecision noise associated with the spectator's estimate of the environmental noise $\xi_{\rm s}(t)$.  Finding the optimal scaling of $\LamImp(t)$ with photon number is thus analogous to minimizing a parameter estimation error with photon number, a standard task in quantum metrology.  We make this connection explicit in what follows. 
We show that in the long-time limit, given a total photon number $\nbartot$ incident on the signal port of the homodyne detector beamsplitter, the measurement imprecision noise dephasing scales as $\LamImp(t)\sim1/\nbartot$ in the case where $\lambda_2 = 0$ (no squeezing).
This is standard quantum limit (SQL) scaling with photon number, expected in the absence of squeezing or entanglement. We also show that if one uses an optimal parametric drive $\lambda_2$, one can improve this scaling to $\LamImp(t)\sim1/\nbartot^2$.  This corresponds to Heisenberg limited scaling with photon number.  
In addition to these fundamental long-time scalings, we also discuss below the minimization of $\LamImp(t)$ at finite evolution times $t_0<\infty$, subject to experimentally relevant constraints for $\lambda_1$ and $\lambda_2$.

In the long-time limit, the qubit is only sensitive to the zero frequency environmental noise $\xi_{\rm q}(t)$.  Hence, noise mitigation in this limit requires the spectator mode to effectively estimate the zero-frequency noise it experiences with minimal error, i.e.~estimate the parameter $\xi_{\rm s}[0] \equiv \lim_{T\to\infty} (1/T)\int_{0}^{T} dt\xi_{\rm s}(t)$ of $\xi_{\rm s}(t)$. 
(The choice of a $1/T$ normalization here ensures that the parameter $\xi_{\rm s}[0]$ being estimated is infinitesimally small.  This lets us connect to standard quantum metrology limits, cf.~Methods Sec.~\ref{app:parameter-estimation}.)
As we show below, the measurement imprecision noise dephasing in this limit, $\LamImp(t\to\infty)$, is directly proportional to the estimation error of $\xi_{\rm s}[0]$.

We can make a direct analogy to the problem of optimally estimating a small phase space rotation $\theta\ll1$ of a single-mode squeezed displaced state.
For our setup in the long time limit, the relevant mode is the zero-frequency output field mode of the spectator, and the rotation $\theta$ is created by zero frequency environmental noise $\xi_{\rm s}[0]$ (see  Fig.~\ref{fig:aout-blob}(a)). 
The results of this basic, single-mode parameter estimation problem are reviewed in Methods Sec.~\ref{app:parameter-estimation}.  The optimal measurement is a homodyne measurement of the phase quadrature of the state, and Heisenberg scaling of the estimation error with photon number is achieved by balancing the number of squeezing photons and displacement photons.  

We now make the analogy to the single-mode parameter estimation problem precise.  The optimal estimator for the zero-frequency environmental noise $\xi_{\rm s}[0]$ will be proportional to the average of the integrated output-field phase quadrature (which is also proportional to our homodyne current).
We thus introduce the zero-frequency output-field temporal mode $\hat{A}$:
\begin{align}
    \hat{A} &= \frac{1}{\sqrt{T}}\int_0^{T} dt\,\hat{a}^{\rm out}(t),
    \label{eq:temporal-mode}
\end{align}
where we will consider the large-$T$ limit throughout. 
This is a standard bosonic mode satisfying $[\hat{A},\hat{A}^\dagger]=1$.
The expectation value of its phase quadrature [$\hat{P}_A = {\rm i}(\hat{A}^\dagger-\hat{A})/\sqrt{2}$] is
\begin{align}
    \langle\hat{P}_A\rangle = \frac{2\detr \sqrt{2\nbardi \kappa_{\rm c} T} }{1+\lambda_{2}} \frac{\xi_{\rm s}[0]}{\kappa_{\rm c}}.
\end{align}
Hence, up to a prefactor, $\hat{P}_A$ will be our estimator for $\xi_{\rm s}[0]$, and the fluctuations in this quadrature $(\Delta P_A)^2 \equiv \langle\hat{P}_A^2\rangle - \langle\hat{P}_A\rangle^2$ will determine our estimation error.  In the long-time limit, this variance just reflects the squeezing created by the parametric drive:
\begin{align}
    (\Delta P_A)^2 &= \frac{(1-\lambda_{2})^{2}}{(1+\lambda_{2})^{2}}.
\end{align}
The estimation error $\Delta\xi_{\rm s}[0]$ of the zero frequency component $\xi_{\rm s}[0]$ is thus given by
\begin{align}
    \Delta \xi_{\rm s}[0] &= \frac{\Delta P_A}{\left| \partial\langle\hat{P}_A\rangle/\partial\xi_{\rm s}[0] \right|}
    = \frac{(1-\lambda_2)\kappa_{\rm c}}{2\detr\sqrt{2\nbardi \kappa_{\rm c} T}}.
    \label{eq:EstimationError}
\end{align}

We would like to understand how this estimation error scales with the number of photons $\nbartot$ used to make the estimate.  This is just the average photon number of the $\hat{A}$ mode, which also coincides with the number of photons incident on our homodyne detector during the measurement interval $T$.  We have: 
\begin{align}
    \nbartot &= \langle\hat{A}^{\dagger}\hat{A}\rangle = \nbard + \nbarsq.
\end{align}
where
\begin{align}
    \nbard = \frac{\lambda_1^2}{(1-\lambda_2)^2} \kappa_{\rm c} T, 
    \qquad
    \nbarsq = \frac{4\lambda_2^2}{(1-\lambda_2^2)^2}. \label{eq:photon-numbers}
\end{align}
The two terms here correspond to photons $\nbard = |\langle\hat{A}\rangle|^2$ associated with the displacement of the mode, and photons $\nbarsq$ associated with the squeezing of the mode.  
Note that $\nbard = \nbardi\kappa_{\rm c} T$ 
(cf.~Eq.~\eqref{eq:displacement-nbar}).  This is just the average output photon flux induced by  $\lambda_1$, $\nbardi \kappa_{\rm c}$, integrated over a time $T$.
Note also that $\nbarsq$ is independent of $T$ because squeezing produces a broadband $(\sim\kappa_{\rm c})$ output photon flux, but the temporal mode has a narrow bandwidth $\sim 1/T \ll \kappa_{\rm c}$ over which it admits squeezing photons; one may show that the $\propto 1/T$ fraction of a $\propto T$ total number of output squeezing photons is thus constant in the long-time limit.
We see immediately that if we do not use any squeezing (i.e., $\lambda_2 = 0$), then $\nbartot = \nbard$, and the estimation error in Eq.~\eqref{eq:EstimationError} exhibits the expected SQL scaling with photon number,  $\Delta\xi_{\rm s}[0] \propto 1 / \sqrt{\nbartot}$.  

To see how the squeezing induced by a non-zero parametric drive $\lambda_2$ can help us, we can re-write the above expression in terms of the photon numbers $\nbard$ and $\nbarsq$.  In the limit where all photon numbers $\nbartot,\nbard,\nbarsq\gg1$, we have $(1-\lambda_2) = 1/\sqrt{\nbarsq} + \mathcal{O}(1/\nbarsq)$, and the estimation error reduces to:
\begin{align}
    \Delta \xi_{\rm s}[0] &= \frac{\kappa_{\rm c}}{2\detr\sqrt{2\nbard\nbarsq}}.
\end{align}

The final step is to now fix the total incident photon number $\nbartot$, and to minimize the estimation error over how we partition these photons between squeezing photons $\nbarsq$ and displacement photons $\nbard$.  In complete analogy to the analysis in Methods Sec.~\ref{app:parameter-estimation}, we find that the optimal partition is an equal split, $\nbard = \nbarsq = \nbartot/2$.  The resulting optimized estimation error is then
\begin{align}
    \Delta \xi_{\rm s}[0] &= \frac{\sqrt{2}\kappa_{\rm c}}{\detr\nbartot}.
    \label{eq:est-error}
\end{align}
We obtain a scaling $1/\nbartot$, which is Heisenberg-limited scaling in the number of photons used for the estimation \cite{giovannetti_QuantumEnhanced_2004}.

The last step here is to show that the estimation error does indeed coincide with the measurement imprecision noise qubit dephasing $\LamImp(T)$ in the long time limit.  
We focus on the optimal tuning of the transduction factor defined in Eq.~\eqref{eq:alpha-s} for noise mitigation, $\as = 1$ (something that can be achieved for any photon number by tuning the feedforward rate $\gff$). 
The interpretation of $\LamImp(T)$ as the estimation error only makes sense for optimal $\as = 1$.
Using Eq.~\eqref{eq:LambdaAddLimits}, we find that in this limit, we have (for arbitrary $\nbard$ and $\nbarsq$):
\begin{align}
    \LamImp(T) &= \frac{1}{4}T^2 
    \left( \Delta \xi_{\rm s}[0] \right)^2.
\end{align}
We see that the spectator-added noise has the same dependence on photon numbers as the estimation error.  Hence, optimizing the estimation error at fixed $\nbartot$ is also optimizes $\LamImp(T)$, and the Heisenberg-limit scaling $ \left( \Delta \xi_{\rm s}[0] \right)^2 \propto 1 / \nbartot^2 $ is also inherited by $\LamImp(T)$.  
(The apparent $T^2$-dependence of $\LamImp$ here may seem strange. However, this is simply an artifact of writing $\LamImp$ in terms of $\nbard$ which is implicitly $\propto T$. Holding $\nbard \propto \lambda_1^2 \kappa_{\rm c} T$ fixed while increasing $T$ requires reducing the linear drive strength $\lambda_1$ in proportion.)

The potential for Heisenberg-limited scaling is interesting from the perspective of fundamental performance limits.
In App.~\ref{app:internal-loss} and Fig.~\ref{fig:internal-loss}, we discuss the breakdown of this Heisenberg-limited scaling caused by non-zero internal loss in the spectator mode.
While these fundamental scaling constraints involve output photon number, from a practical perspective, a more common constraint comes from only being able to work with a finite number of total \emph{intracavity} photons.  These technical limitations can arise for a variety of reasons, e.g., nonlinear effects, heating, input power limits, or a breakdown of the dispersive approximation.
In addition, minimizing $\LamImp(t)$ at some finite time $t_0$ is more important in most cases than minimizing the asymptotic dephasing rate in the long-time limit.

We minimize $\LamImp(t_0)$ (cf.~Eq.~\eqref{eq:lambda-add}) at a fixed target evolution time $t_0<\infty$ over the partition of intracavity photon number $\nbartoti = \nbardi + \nbarsqi$, while holding $\nbartoti \gg 1$ fixed. (The assumption of $\nbartoti\gg1$ is used to identify the long-time regime in what follows; however, $\nbartoti\gtrsim 10$ is sufficient.)
The intracavity displacement photon number $\nbardi = |\langle\hat{a}\rangle|^2$ is given by Eq.~\eqref{eq:displacement-nbar}, and the intracavity squeezing photon number $\nbarsqi$ is given by
\begin{equation}
    \nbarsqi = \frac{1}{2}\frac{\lambda_{2}^{2}}{1-\lambda_{2}^{2}}. \label{eq:squeezing-nbar}
\end{equation}
Recall that $\kappa_\phi = (1+\lambda_2)\kappa_{\rm c}$ is dependent on $\nbarsqi$ through $\lambda_2$, thus it is allowed to vary when minimizing $\LamImp(t_0)$. 
We find that there are three distinct optimization regimes in $t_0$: the short-time regime $t_0\lesssim 1/\kappa_{\rm c}$, the long-time regime $t_0 \gg \nbartoti^2 /\kappa_{\rm c}$, and the intermediate regime $1/\kappa_{\rm c} \lesssim t_0 \lesssim \nbartoti^2/\kappa_{\rm c}$.

In the extreme short-time limit $t_0\ll 1/\kappa_{\rm c}$, squeezing has no ability to reduce $\LamImp(t_0)$   (cf.~Eq.~\eqref{eq:LambdaAddLimits}).
Even for $t_0 \lesssim 1/\kappa_{\rm c}$, squeezing has very little ability to reduce the added-noise dephasing because the qubit is sensitive to noise with a bandwidth $1/t_0 \gtrsim \kappa_{\rm c}$ but the squeezing bandwidth is only $\kappa_{\rm c}$.
Thus, since $\LamImp(t_0) \propto 1/\nbardi$, the optimal partition is $\nbardi \approx \nbartoti$ and $\nbarsqi\approx 0$.

The long-time regime is identified by $t_0$ (already assumed $\gg 1/\kappa_{\rm c}$) being sufficiently large such that the constant term of $\LamImp(t_0)$ is negligible compared to the linear-in-$t_0$ term in Eq.~\eqref{eq:LambdaAddLimits}.
Thus holding $\nbartoti\gg1$ fixed, we assume that $\nbarsqi\gg1$ and write $\LamImp(t)$ to leading order in $1/\nbarsqi$. (The assumption that the optimal partition has $\nbarsqi\gg1$ can be shown to be self-consistent using the exact expression for $\lambda_2$ in terms of $\nbarsqi$.)
\begin{align}
    \LamImp(t_0) = \frac{\as^2}{32\detr^2}\frac{1}{\nbardi}\left( \frac{1}{16\nbarsqi^2}\kappa_{\rm c} t_0 + 4 \right).
\end{align}
From this we find that the constant is negligible when $t_0 \gg \nbarsqi^2/\kappa_{\rm c} \sim \nbartoti^2/\kappa_{\rm c}$.
This regime has the optimal partition
\begin{align}
    \nbarsqi = \frac{2}{3}\nbartoti,~~\nbardi = \frac{1}{3}\nbartoti;\quad(t_0 \gg \nbartoti^2/\kappa_{\rm c}).
    \label{eq:long-time-optimal-frac}
\end{align}
This optimal partition of \emph{intracavity photons} is equivalent to the partition of photons incident on the homodyne detector ($\nbard = \nbarsq = \nbartot/2$ as discussed in the previous subsection) that optimizes the Heisenberg-limited scaling of the measurement error. 
This equivalence is easily shown via Eq.~\eqref{eq:photon-numbers} for a target evolution time $t_0 = T$.

 \begin{figure}[t]
     \centering
    \includegraphics[width=0.999\columnwidth]{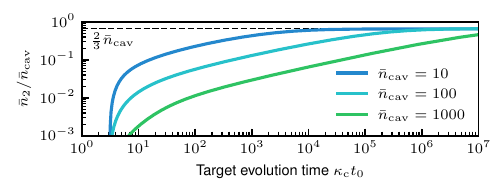}
     \caption{
     \textbf{Optimized intracavity squeezing.}
     The optimal fraction of $\nbartoti$ intracavity photons applied to squeezing, $\nbarsqi/\nbartoti$, vs. target evolution time $\kappa_{\rm c} t_0$ (i.e., the time $t_0$ at which $\LamImp(t_0)$ is minimized) for various $\nbartoti$. 
     For very long times $\kappa_\phi t_0 \gg \nbartoti^3$, the squeezing photon number fraction approaches the asymptotic optimal value $\frac{2}{3}\nbartoti$ (cf. Eq.~\eqref{eq:long-time-optimal-frac}).
     The parameters are $\as = 1$ and $\detr = 1$.}
     \label{fig:opt-intracavity-squeezing}
 \end{figure} 

The cross-over behavior of optimal $\nbarsqi$ between the limiting regimes of $t_0\lesssim 1/\kappa_{\rm c} $ and $t_0 \gg \nbartoti^3/\kappa_{\rm c} $ is shown in Fig.~\ref{fig:opt-intracavity-squeezing}. 
The surprisingly small fraction of photons allocated to squeezing in the intermediate regime might suggest that the use of squeezing is not very effective at reducing the qubit dephasing rate.
However, this is not necessarily the case. 
In Fig.~\ref{fig:intro}(b) we show that for $\kappa_{\rm c} t_0 \sim 1$, a relatively small number of intracavity squeezing photons can still dramatically reduce the long-time qubit dephasing rate.
For the optimal $\lambda_2$ curve in Fig.~\ref{fig:intro}(b), we minimize $\LamImp(t_0)$ at time $t_0 = 5/\kappa_{\rm c}$ ($t_0 = 0.005/S_0 $) to achieve a $\approx 10^{-2}$ reduction in the long-time dephasing rate versus  $\Gamma_0 = S_0/2$.
Using $\nbartoti = 1000$, at $t_0 = 5/\kappa_{\rm c}$, the optimal number of intracavity squeezing photons is only $\nbarsqi^\star \approx 0.6$.


\section{Discussion}\label{sec:conclusion}

Given the range of operating regimes we have considered and the many control parameters one may vary, we briefly summarize how the parameters must be constrained in order to successfully operate the spectator mode.
There are five control parameters: the noise coupling parameter $\detr$ (cf. Eq.~\eqref{eq:Hspec}), the spectator mode decay rate $\kappa_{\rm c}$, the intracavity displacement photon number $\nbardi$ (cf. Eq.~\eqref{eq:displacement-nbar}, via linear drive strength $\lambda_1$), the intracavity squeezing photon number $\nbarsqi$ (cf. Eq.~\eqref{eq:squeezing-nbar}, via parametric drive strength $\lambda_2$), and the feedforward rate to the qubit $\gff$ (cf. Eq.~\eqref{eq:L-diss}).
Successful operation of the spectator mode requires that these parameters be set by the conditions summarized in Table~\ref{tab:parameter}.
If the first two conditions are satisfied, then the long-time dephasing is given in general by Eq.~\eqref{eq:long-time-decoherence-partial-correlation}, where the long-time exponential decay of coherence can be made arbitrarily small in the limit of perfect noise correlation $\eta=1$. 
Condition 3 ensures that the spectator mode is fast enough that the short-time decoherence (i.e., due to the noise outside the spectator mode's bandwidth) is tolerably small.
Finally, condition 4 is a practical constraint: one must balance the partition of photons inside the spectator mode such that at some finite target time $t_0$, the total qubit coherence is minimized.

\begin{table*}
    \caption{Summary of spectator mode operating parameters.}
    \label{tab:parameter}
    \centering
    \begin{tabular}{|p{14 em}|p{24 em}|p{15 em}|}
    \hline
        \textbf{Condition} & \textbf{Purpose} & \textbf{Parameters}\\
        \hline\hline
        Either Eq.~\eqref{eq:linear-valid} or Eq.~\eqref{eq:squeezing-condition} (without or with squeezing) & Effective linear noise coupling dominates spurious nonlinear coupling & $\detr$,~$\kappa_c$ \\
        \hline
        $\as = \as^{\rm ideal} = \eta$ (cf. Eqs.~\eqref{eq:alpha-s} and \eqref{eq:etaDefinition}) & Transduction factor set to ideal value set by noise correlation & $\detr$, $\kappa_c$, $\gff$, $\nbardi$ (via $\lambda_1$), $\nbarsqi$ (via $\lambda_2$)\\
        \hline
        $\chi_{\rm init}(\infty) \ll 1$ (cf.~Eq.~\eqref{eq:chi-init}) & Short time dephasing due to finite cavity response time is tolerably small & $\kappa_c$\\
        \hline
        $\LamImp(t_0)$ minimized for fixed $\nbartoti=\nbardi+\nbarsqi$ (cf. Eq.~\eqref{eq:lambda-add}) & Intracavity photons are partitioned to minimize qubit dephasing at target time $t_0$ & $\nbardi$, $\nbarsqi$\\
        \hline
    \end{tabular}
\end{table*}

We have shown that a spectator photonic mode can harness spatial correlations in classical dephasing noise as a resource for noise mitigation and is a powerful generalization of the spectator qubits concept. 
The spectator mode can perfectly cancel qubit dephasing due to $\xi(t)$ in the long-time limit, and despite having a finite noise detection bandwidth, can mitigate white noise.
The multi-leveled nature of the spectator mode allows the use of many photons to be simultaneously measured, thereby reducing the measurement imprecision noise to arbitrarily small levels.
The use of many photons in the measurement is a significant advantage over spectator qubits which suffer the imprecision error of single measurements.
The spectator mode even offers better than standard quantum limit scaling of the measurement imprecision error with the number of photons: using parametric driving, we show that it achieves Heisenberg-limited scaling.
Finally, we show that even under the constraint of a limited intracavity photon number, the use of a parametric drive dramatically improves the performance of the spectator mode, even if true Heisenberg-limited scaling is not achieved.

The spectator mode scheme is readily amenable to experimental implementations; superconducting circuits, for example, provide a natural experimental platform in which to implement the scheme and experimentally verify these results.
Not only is this platform among the leading candidates for large scale quantum information processing, all of the necessary ingredients already exist and are frequently used in superconducting circuit experiments, including frequency tunable qubits with fast flux control \cite{gargiulo_Fast_2021,zhang_Universal_2021}, photonic (microwave) cavities that can be dispersively coupled to a source of correlated noise, and homodyne detection of the cavity output field.
The full noise environment of a typical superconducting qubit is not yet fully understood and many sources of noise may contribute to qubit decoherence.
For example, flux noise is a common decoherence mechanism in flux-tunable qubits whose microscopic origins are not completely understood yet, although the experiment in Ref.~\cite{kou_FluxoniumBased_2017} empirically finds no strong spatial correlations.
Nevertheless, there are many sources of environmental noise that the spectator mode could detect and mitigate.

In quantum registers with many qubits and their associated control line, electromagnetic crosstalk from nearby qubits and control lines appears as dephasing noise \cite{sarovar_Detecting_2020,tripathi_Suppression_2022}; this noise is spatially correlated essentially by definition.
Thus, by equipping the spectator mode with, e.g., a SQUID loop, it can dispersively detect magnetic interference from neighboring qubits and their control lines.
In the same vein, residual ZZ couplings between physically proximate qubits can induce decoherence in the target data qubit if a neighboring qubit undergoes a T1 relaxation, thereby effectively applying a phase kick to the data target data qubit \cite{jurcevic_Effective_2022}.
In principle, by weakly dispersively coupling the spectator mode to the nearby qubits, it can continuously detect these relaxation events and apply corrective phase kicks.
As a final example, it is well known that when qubits are dispersively coupled to a readout resonator, residual photon population (e.g., thermal population) in the readout resonator can dephase the qubits \cite{rigetti_Superconducting_2012,gambetta_Qubitphoton_2006}.
The fluctuations in the readout resonator photon number cause small changes in the qubits' spitting frequencies, thereby dephasing the qubits.
A spectator mode dispersively coupled to the readout resonator could detect these photon number fluctuations and apply a corrective signal to the qubits.

In this work we have explored the basic spectator mode system.
We use a linear measurement of the phase quadrature which necessitates neglecting higher order corrections due to spectator mode dephasing.
A more sophisticated measurement apparatus could improve upon that, and perhaps relax the requirement for sufficiently weak noise.
Recent work in using various adaptive algorithms \cite{wiseman_Quantum_2009,gupta_Adaptive_2020} or machine learning \cite{gupta_Machine_2018,mavadia_Prediction_2017} has been shown to improve the performance spectator qubits. 
It remains an open question whether the use of such techniques could improve the performance of spectator modes and to what extent there are optimal strategies for specific spectral densities or noise models (e.g., $1/f$, telegraph, etc).
Similarly, the study of more general noise models (i.e., going beyond classical stationary noise) is an open direction for future work, not only within the context of the spectator photonic mode presented here, but more broadly within the entire paradigm of spectator quantum systems.

Furthermore, the generalization of the spectator qubit concept could lead to other spectator quantum systems such as qubit or spin ensembles.
The recent experiment reported in Ref.~\cite{singh_Midcircuit_2023} has many spectator qubits available; using those in a coordinated way may offer some of the advantages of the spectator mode regarding measurement imprecision.
Furthermore, one could imagine that the advantages of parametric driving to the spectator mode could be realized in spin ensembles using spin squeezing.
These are all additional avenues of the spectator quantum system concept to explore.


\section{Methods}

\subsection{Derivation of accumulated phase and measurement imprecision noise dephasing}
\label{app:decoherence-func}

In this Section we derive the accumulated phase $\phi_{[\xi]}(t)$ and measurement imprecision noise dephasing $\LamImp(t)$ starting from the master equation for $\hat{\rho}_{[\xi]}(t)$, Eq.~\eqref{eq:qme}. 
For the analysis in the next Subsection on the linear noise drive approximation, we retain the spurious phase noise term in Eq.~\eqref{eq:Hspecprime}:
\begin{align}
    \hat{H}_{{\rm spec},[\xi]}^\prime(t) &=\epsilon \detr \xi_{\rm s}(t)\hat{d}^{\dagger}\hat{d}+\detr\xi_{\rm s}(t)\sqrt{\nbardi}(\hat{d}+\hat{d}^{\dagger}) \label{eq:Hs-phase-noise}\\
    &-\frac{{\rm i}\kappa_{\rm c}\lambda_{2}}{4}(\hat{d}^{2}-\hat{d}^{\dagger2}). \nonumber
\end{align}
To differentiate phase noise terms from the quadrature noise, we include a bookkeeping prefactor $\epsilon$ in the phase noise terms. 
The full Hamiltonian has $\epsilon=1$ and the linear noise drive approximation Eq.~\eqref{eq:Hs} has $\epsilon=0$. 

In writing Eq.~\eqref{eq:qme} for $\hat{\rho}_{[\xi]}(t)$, we have assumed that the measurement and feedforward is running at all times. 
Instead, the anticipated operation would have the feedforward turned off and the spectator mode in steady state at the start of the experiment.
At a definite time $t=0$, i.e., the start of the experiment, the qubit has a known coherence $\rho_{\uparrow\downarrow}(0)$, thus we neglect all qubit dynamics for $t<0$ by modifying Eq.~\eqref{eq:Hq} to
\begin{align}
    \hat{H}_{{\rm q},[\xi]}(t)&=\frac{1}{2}\xi_{\rm q}(t)\hat{\sigma}_{\rm z}\theta(t),
\end{align}
where $\theta(t)$ is the step function. To implement the turn-on of the feedforward in the master equation at $t=0$,  we modify Eqs.~\eqref{eq:Hint} and \eqref{eq:L-diss} 
\begin{align}
    \hat{H}_{\mathrm{int}}(t)&=\frac{1}{2{\rm i}}\sqrt{\gff\kappa_{\rm c}}\left(\hat{d}-\hat{d}^{\dagger}\right)\hat{\sigma}_{\rm z}\theta(t),\\
    \hat{L}(t)&=\sqrt{\kappa_{\rm c}}\hat{d}+\sqrt{\gff}\hat{\sigma}_{\rm z}\theta(t).
\end{align}
For $t<0$, the master equation simply describes a damped, driven oscillator dispersively coupled to the environmental noise.

With the above refinement in hand, we now find the deformed master equation of the coherence operator \cite{clerk_Using_2007}, 
\begin{align}
    \hat{\rho}_{\uparrow\downarrow,[\xi]}(t)\equiv\langle\uparrow\rvert\hat{\rho}_{[\xi]}(t)\lvert\downarrow\rangle.
\end{align}
The equation of motion for the coherence operator is (dropping its explicit $t$-dependence and $[\xi]$ subscript for clarity)
\begin{align}
    \frac{d}{dt}&\hat{\rho}_{\uparrow\downarrow}=-{\rm i}[\hat{H}_{\rm s},\hat{\rho}_{\uparrow\downarrow}] +\kappa_{\rm c}\mathcal{D}[\hat{d}]\hat{\rho}_{\uparrow\downarrow}\\
    &-\theta(t)\left[2\sqrt{\kappa_{\rm c}\gff}(\hat{d}\hat{\rho}_{\uparrow\downarrow}-\hat{\rho}_{\uparrow\downarrow}\hat{d}^{\dagger}) + 2\gff\hat{\rho}_{\uparrow\downarrow}-{\rm i}\xi_{\rm q}(t)\hat{\rho}_{\uparrow\downarrow}\right] \nonumber
\end{align}
We represent $\hat{\rho}_{\uparrow\downarrow,[\xi]}(t)$  by its Wigner function $W_{[\xi]}(x,p;t)$ via
\begin{align}
    W_{[\xi]}(x,p;t)&=\frac{1}{\pi}\int dy{\rm e}^{-2{\rm i}py}\langle x+y|\hat{\rho}_{\uparrow\downarrow,[\xi]}(t)|x-y\rangle
\end{align}
where $|x\rangle$ are the amplitude quadrature eigenstates, $\hat{X}|x\rangle=x|x\rangle$. We thus derive the equation of motion for $W_{[\xi]}(x,p;t)$:
\begin{widetext}
\begin{align}
    \partial_{t}W_{[\xi]}(x,p;t)=&\left[\epsilon \detr\xi_{\rm s}(t)\left(x\partial_{p}-p\partial_{x}\right)+\detr\xi_{\rm s}(t)\sqrt{2\nbardi}\partial_{p}+\frac{\kappa_{\rm c}}{2}\left(\partial_{x}x+\partial_{p}p + \frac{1}{2}\partial_{p}^{2}+\frac{1}{2}\partial_{x}^{2}\right) \right.\label{eq:Wigner-eom}\\
    &~~\left.\vphantom{\int} + \frac{\kappa_{\rm c}}{2}\lambda_{2}\left(\partial_{p}p-\partial_{x}x\right) - i\xi_{\rm q}(t)\theta(t)-{\rm i}\sqrt{2\kappa_{\rm c}\gff}\left(\partial_{p}+2p\right)\theta(t)-2\gff\theta(t)\right]W_{[\xi]}(x,p;t). \nonumber
\end{align}
\end{widetext}
This deformed Fokker-Planck equation is at most quadratic in $x$, $p$, and their derivatives, as expected because the original master equation is at most quadratic in $\hat{d},\hat{d}^\dagger$. 
Thus for a Gaussian initial condition of the spectator, the Wigner function is Gaussian for all times. 

We introduce the Fourier-transformed  Wigner function via
\begin{align}
    W(x,p;t)=\int\frac{dk}{2\pi}\int\frac{dq}{2\pi}{\rm e}^{{\rm i}kx+{\rm i}qp}W[k,q;t],
\end{align}
for which the Gaussian ansatz is given by
\begin{align}
    \frac{W[k,q;t]}{W[0,0;0]}&=\exp\left[-{\rm i}\nu(t)+{\rm i}k\bar{x}(t)+{\rm i}q\bar{p}(t)\right] \label{eq:Gaussian-ansatz}\\
    &\times\exp\left[-\frac{1}{2}\{k^{2}\sigma_{x}(t)+q^{2}\sigma_{p}(t)\}-kq\sigma_{xp}(t)\right]. \nonumber
\end{align}
Here the Gaussian parameters are: the quadrature means $\bar{x}(t),\bar{p}(t)$, the covariances $\sigma_{x}(t)$, $\sigma_{p}(t)$, and $\sigma_{xp}(t)$, and finally the overall phase $\nu(t)$ which yields the stochastic qubit coherence:
\begin{align}
    \rho_{\uparrow\downarrow,[\xi]}(t)=\mathrm{tr}_{\rm s}[\hat{\rho}_{\uparrow\downarrow,[\xi]}(t)]=\int dx dpW_{[\xi]}(x,p;t)={\rm e}^{-{\rm i}\nu(t)}.
\end{align}

In general $\nu(t)$ is complex-valued; the real part is a true accumulated phase, and the imaginary part is the mean loss of coherence due to the quantum average over measurement outcomes implicit in the unconditional master equation. 
We thus identify the real and imaginary parts of $\nu(t)$ with the accumulated phase $\phi_{[\xi]}(t)$ and the measurement imprecision noise decoherence $\LamImp(t)$ (cf. Eqs.\eqref{eq:phi} and \eqref{eq:lambda-add}):
\begin{align}
    \phi_{[\xi]}(t)&\equiv\mathrm{Re}[\nu(t)];\quad
    \LamImp(t)\equiv -\mathrm{Im}[\nu(t)].
\end{align}

After substituting the Gaussian ansatz in Eq.~\eqref{eq:Wigner-eom} and matching Fourier coefficients, we find
\begin{align}
    \partial_{t}\nu(t)=&\left[-\xi_{\rm q}(t) + 2\sqrt{2\kappa_{\rm c}\gff}\bar{p}(t) +2{\rm i}\gff\right]\theta(t),
    \label{eq:nu-general}\\
    \partial_{t}\bar{x}(t)=&-\frac{\kappa_{\rm a}}{2}\bar{x}(t)+\epsilon \detr\xi_{\rm s}(t)\bar{p}(t)\\
    &+2{\rm i}\sqrt{2\kappa_{\rm c}\gff}\sigma_{xp}(t)\theta(t), \nonumber\\
    \partial_{t}\bar{p}(t)=&-\frac{\kappa_{\phi}}{2}\bar{p}(t) -\epsilon \detr\xi_{\rm s}(t)\bar{x}(t) + \sqrt{2\nbardi}\detr\xi_{\rm s}(t) \\
    &+{\rm i}\sqrt{2\kappa_{\rm c}\gff}\left[2\sigma_{p}(t) - 1\right]\theta(t), \nonumber\\
    \partial_{t}\sigma_{x}(t)=&-\kappa_{\rm a}\sigma_{x}(t)+\frac{\kappa_{\rm c}}{2} +2\epsilon \detr\xi_{\rm s}(t)\sigma_{xp}(t),\\
    \partial_{t}\sigma_{p}(t)=&-\kappa_{\phi}\sigma_{p}(t)+\frac{\kappa_{\rm c}}{2} -2\epsilon \detr\xi_{\rm s}(t)\sigma_{xp}(t),\\
    \partial_{t}\sigma_{xp}(t)=&-\kappa_{\rm c}\sigma_{xp}(t)+\epsilon \detr\xi_{\rm s}(t)\left[\sigma_{p}(t)-\sigma_{x}(t)\right],\label{eq:sigma-xp-general}
\end{align}
where $\kappa_{\phi} = \kappa_{\rm c}(1+\lambda_2)$ (cf. Eq.~\eqref{eq:kappa-phi}) and $\kappa_{\rm a}=\kappa_{\rm c}(1-\lambda_{2})$ is the amplitude quadrature damping rate. 
The overall phase $\nu(t)$ has three contributions: the direct environmental noise $\xi_{\rm q}(t)$, the added noise from the homodyne current $\propto\gff$, and the signal in the homodyne current $\propto\bar{p}(t)$ containing both the environmental and added quantum noises.

Making the linear noise drive approximation (by setting $\epsilon=0$), we find the overall phase $\nu(t)$:
\begin{align}
    \nu&(t)= \label{eq:linear-nu-expression}\\
    &\int_{0}^{t}dt^{\prime}\left[\xi_{\rm q}(t^{\prime}) -  \frac{\kappa_{\phi}}{2}\as\int_{-\infty}^{t^{\prime}}ds{\rm e}^{-\kappa_{\phi}(t^{\prime}-s)/2}\xi_{\rm s}(s)\right] \nonumber\\
    &-\frac{2{\rm i}\gff}{(1+\lambda_{2})^{2}}\left[(1-\lambda_{2})^{2}t+\frac{8\lambda_{2}}{\kappa_{\phi}}\left(1-{\rm e}^{-\kappa_{\phi}t/2}\right)\right]. \nonumber
\end{align}
We immediately read off the real part of $\nu(t)$ as  Eq.~\eqref{eq:phi}, and the imaginary part of $\nu(t)$ as Eq.~\eqref{eq:lambda-add}, upon replacement of $\gff$ using Eq.~\eqref{eq:alpha-s}.


\subsection{Linear noise drive approximation}
\label{app:linearization}

We analyze here the effects of the unwanted displaced-frame spurious phase noise term  $\detr\xi_{\rm s}(t)\hat{d}^\dagger\hat{d}$ (c.f.~Eq.~\eqref{eq:Hspecprime}), and derive Eqs.~\eqref{eq:linear-valid} and \eqref{eq:squeezing-condition} 
that determine when it is valid to drop this interaction.  We also derive an approximate expression that describes the small amount of extra qubit dephasing that would result from retaining the spurious phase noise coupling, and confirm via explicit numerical simulations its validity.

We start with Eqs.~\eqref{eq:nu-general}-\eqref{eq:sigma-xp-general} with $\epsilon=1$, so as to retain the  phase noise terms.
Generically, these equations are not analytically solvable for nonzero squeezing ($\lambda_{2}\neq0$).
We discuss the interplay of phase noise and squeezing below.
For now, we let $\lambda_{2}=0$ which sets $\kappa_{\phi}=\kappa_{\rm a}=\kappa_{\rm c}$ in Eqs.~\eqref{eq:nu-general}-\eqref{eq:sigma-xp-general}. 
With one decay rate for both quadratures, the equations are solvable. 
The overall complex qubit phase is thus given by
\begin{align}
    &\nu(t)= -2{\rm i}\gff t \label{eq:nonlinear-nu-expression}\\
    &+\int_{0}^{t}\!\!dt^{\prime} \left[ \xi_{\rm q}(t^{\prime}) - \frac{\kappa_{\rm c}}{2}\as \int_{-\infty}^{t^{\prime}}\!\!\!\!\!\! ds {\rm e}^{-\frac{\kappa_{\rm c}}{2}(t^{\prime}-s)} \xi_{\rm s}(s) \cos\psi(t^{\prime},s) \right] \nonumber
\end{align}
where $\psi(t^{\prime},s)$ is the accumulated phase due to the
displaced-frame spurious phase noise:
\begin{align}
    \psi(t^{\prime},s)\equiv\epsilon \detr\int_{s}^{t^{\prime}}d\tau\xi_{\rm s}(\tau).
\end{align}
From this we see that making the linear noise drive approximation (setting $\epsilon=0$) amounts to assuming $\psi(t,s)=0$ (cf.~the real part of Eq.~\eqref{eq:nonlinear-nu-expression} vs. the real part of Eq.~\eqref{eq:linear-nu-expression}).

This exact closed-form expression for the accumulated phase $\phi_{[\xi]}(t)=\mathrm{Re}[\nu(t)]$ is a remarkably simple extension of Eq.~\eqref{eq:phi}.  Unfortunately, it makes $\phi_{[\xi]}(t)$ a  nonlinear function of $\xi_{\rm s}(t)$, making analytic classical stochastic averaging challenging.  While 
perturbative approaches to this problem are possible (see e.g., Refs.~\cite{zhang_Interplay_2014,zhang_Spectral_2015}), we consider a more heuristic  approach.

We want to find conditions where it is safe to neglect the unwanted frequency noise, i.e., approximate the phase $\psi(t',s) \simeq 0$.  Note crucially that from Eq.~\eqref{eq:nonlinear-nu-expression}, contributions from this phase are exponentially suppressed if $t'-s > 1 / \kappa_{\rm c}$:  the cavity only experiences this phase diffusion over a limited time window $\sim 1 / \kappa_{\rm c}$.  This observation then motivates the following heuristic condition:  the effect of frequency noise (in the displaced frame) can be neglected as long as the RMS value of $\psi(t',s) $ is small on timescales $t'-s \sim 1/\kappa_{\rm c}$, i.e.
\begin{align}
    1 - \overline{\cos\psi(1/\kappa_{\rm c},0)} \approx \frac{1}{2}\overline{\psi(1/\kappa_{\rm c},0)^2} \ll 1
    \label{eq:small-condition}
\end{align}
Computing the variance of $\psi(1/\kappa_{\rm c},0)$ in terms of the noise spectral density, the condition of Eq.~\eqref{eq:small-condition}  becomes 
\begin{align}
    \frac{1}{2}\detr^{2}\int\frac{d\omega}{2\pi}\frac{S[\omega]}{\omega^{2}}|Y_{\rm fid}(\omega,1/\kappa_{\rm c})|^{2} \ll 1,
    \label{eq:lnda-app}
\end{align}
where $Y_{\rm fid}(\omega,\tau)$ is the free induction decay filter function in Eq.~\eqref{eq:fid-filter-func}. 
Substituting for $Y_{\rm fid}(t)$, this becomes Eq.~\eqref{eq:linear-valid}.

Because the expression Eq.~\eqref{eq:phi} for the accumulated qubit phase $\phi_{[\xi]}(t)$ is an approximation, the perfect noise cancellation that can be achieved for optimal $\as = 1$ is only approximately true. 
Here we seek an estimate of the additional small qubit dephasing that will arises from the displaced-frame spurious phase noise, Eq.~\eqref{eq:Hfreq}.  We again focus on the no squeezing case, $\lambda_2=0$, and assume
that the spectral density always satisfies Eq.~\eqref{eq:linear-valid} for a given $\kappa_{\rm c}$.

In Eq.~\eqref{eq:nonlinear-nu-expression}, whose real part is $\phi_{[\xi]}(t)$, the $\propto\as$ term will typically be suppressed by the phase noise.
Following the argument above, a nonzero RMS value of $\psi(1/\kappa_{\rm c},0)$ results in $\overline{\cos\psi(1/\kappa_{\rm c},0)}<1$, thereby effectively renormalizing $\as$ to $\tilde{\alpha}_{\rm s} < \as$.
We equate the renormalized linear-in-$\xi_{\rm s}(t)$ qubit phase accumulation (with transduction strength $\tilde{\alpha}_{\rm s}$) to the full qubit phase accumulation with the RMS-averaged phase $\psi(1/\kappa_{\rm c},0)$:
\begin{align}
    \tilde{\alpha}_{\rm s} \approx \as\overline{\cos\psi(1/\kappa_{\rm c},0)} 
    \label{eq:renormalized-alpha}
\end{align}
This renormalization effect suggests a simple mitigation strategy:  to optimally cancel long-time qubit dephasing, one should increase $\as$ above one, so that the renormalized parameter $\tilde{\alpha}_{\rm s} = 1$.  In practice, this tuning would happen naturally if one adjusted the experimental parameters that control $\as$ to minimize the long-time dephasing.  This suggests that our general strategy can be effective even in regimes where the leading effects of the residual displaced-frame phase noise contribute.  

In what follows, we will use Eq.~\eqref{eq:renormalized-alpha} to test whether our approximate treatment of the spurious phase noise is indeed accurate.  If $\as = 1$, then the renormalization in Eq.~\eqref{eq:renormalized-alpha} will lead to a small amount of extra dephasing.  Using  Eq.~\eqref{eq:long-time-dephasing} the long-time qubit dephasing is now given by
\begin{align}
    \chi(t\to\infty) &= \frac{1}{2}\left[ \detr^2 \int \frac{d\omega}{2\pi}  \frac{S[\omega]}{\omega^2} |Y_{\rm fid}(\omega,1/\kappa_{\rm c})|^2 \right] S[0]t \nonumber \\
    &+ \LamImp(t) + \chi_{\rm init}(\infty)
\end{align}
where the first term is our estimate for the extra dephasing from cavity phase noise (expanded to lowest order in the noise).  We see that this extra dephasing is suppressed over the bare dephasing $\Gamma_0=S[0]/2$ by the LHS of  Eq.~\eqref{eq:linear-valid}:
\begin{align}
    \frac{\Gamma_{\rm res}}{\Gamma_0} \approx \left[\detr^2 \int \frac{d\omega}{2\pi}  \frac{S[\omega]}{\omega^2} |Y_{\rm fid}(\omega,1/\kappa_{\rm c})|^2 \right] \ll 1.
    \label{eq:residual-dephasing-est}
\end{align} 
Next, we will text this approximation against numerically-exact stochastic averaging, showing a good agreement.  This shows that our approximate treatment of the residual displaced-frame phase noise is valid.

To check the validity of the linear noise drive approximation, we numerically compute the qubit coherence function under the full spectator Hamiltonian Eq.~$\eqref{eq:Hspecprime}$ using generated noise time series and averaging over the noise realizations. We numerically integrate Eqs.~\eqref{eq:nu-general}-\eqref{eq:sigma-xp-general} for the Gaussian parameters and neglect the quantum effects (i.e., the imaginary parts).

Neglecting squeezing ($\lambda_2=0$) to simplify the numerical integration, we compare the numerically computed dephasing function $\chi_{\rm n}(t)$ to the analytic expression Eq.~\eqref{eq:dephasing-func} calculated under the linear noise drive approximation.
We consider Lorentzian noise 
\begin{align}
    S[\omega] = S_0 (\gamma^2/4)/(\omega^2 +\gamma^2/4)
\end{align}
of moderate to large bandwidth $\gamma \geq \kappa_{\rm c}$ and varying zero-frequency noise strengths $10^{-3} \leq S_0/\kappa_{\rm c} \leq 10^{-1}$.

\begin{figure}[t!]
    \centering
    \includegraphics[width=0.999\columnwidth]{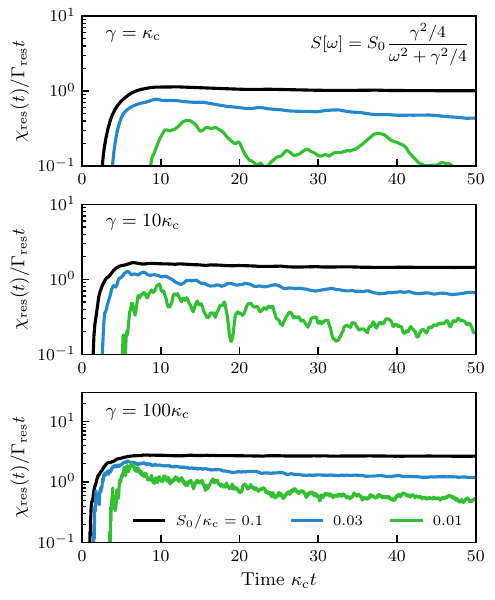}
    \caption{
    \textbf{Numerical verification of the linear noise drive approximation.}
    The ratio of residual dephasing due to the spurious phase noise $\chi_{\rm res}(t)$  (cf. Eq.~\eqref{eq:chi-res}) to the estimated residual linear dephasing (rate $\Gamma_{\rm res}$, cf. Eq.~\eqref{eq:residual-dephasing-est}) vs. time. 
    Here we consider Lorentzian noise of indicated zero-frequency strengths $S_0 = S[0]$ (this is the relevant small parameter which should satisfy $S_0/\kappa_{\rm c}\ll1$).
    Each plot shows a different noise bandwidth $\gamma$, as indicated.
    We see that, after some initial nonlinear residual dephasing, the linear rate estimate Eq.~\eqref{eq:residual-dephasing-est} is a reasonable estimate of the residual dephasing, validating the linear noise drive approximation.
    Here we take $\detr=1$ and $\lambda_2=0$.
    The linear drive strength $\lambda_1$ is irrelevant.}
    \label{fig:linear-noise-drive-verification}
 \end{figure}

We average over $10^5$ noise realizations for each numerically integrated $\chi_{\rm n}(t)$.
We verify convergence of the numerical simulations by also numerically integrating the linear noise drive approximation decoherence functions at each parameter choice and comparing with the analytic expressions.
To compare the approximation with the full spectator dynamics, we compute a residual dephasing function
\begin{align}
    \chi_{\rm res}(t) = \chi_{\rm n}(t) - \chi(t),
    \label{eq:chi-res}
\end{align}
where $\chi(t)$ is the analytic expression Eq.~\eqref{eq:dephasing-func}, neglecting $\LamImp(t)$.
This is compared with the estimated linear residual dephasing with rate $\Gamma_{\rm res}$ given by Eq.~\eqref{eq:residual-dephasing-est}.
The results are shown in Fig.~\ref{fig:linear-noise-drive-verification} where we plot the ratio $\chi_{\rm res}(t)/(\Gamma_{\rm res}t)$ as a function of time.
We see that Eq.~\eqref{eq:residual-dephasing-est} is a good order-of-magnitude estimate of the residual long-time dephasing rate.

The interplay between classical spurious phase noise (c.f.~Eq.~\eqref{eq:Hfreq}) and squeezing places further constraints on the operating regime of the spectator mode beyond Eq.~\eqref{eq:lnda-app}.  This is due to the random mixing of canonical squeezed and amplified quadratures by the phase noise.  
Generically, the Gaussian parameter equations of motion 
(cf.~Eqs.~\eqref{eq:nu-general}-\eqref{eq:sigma-xp-general}) are not solvable in closed form when the spectator mode is parametrically driven, $\lambda_2\neq 0$.
However, following Refs. \cite{zhang_Quantum_2003,aoki_Squeezing_2006,oelker_Ultralow_2016}, we obtain the following heuristic estimate for how much the phase noise degrades the squeezing.

The phase noise generates an RMS average rotation of phase space by angle $\theta_{\rm rms}$, thus rotating the maximally-amplified amplitude quadrature ($X$) fluctuations into the maximally-squeezed phase quadrature ($P$):
\begin{align}
    (\Delta P)^2 = (\Delta P^{(0)})^2\cos^2\theta_{\rm rms} + (\Delta X^{(0)})^2 \sin^2\theta_{\rm rms}
\end{align}
The unperturbed quadrature fluctuation levels are $\Delta P^{(0)} = (1-\lambda_2)/(1+\lambda_2)$ and $\Delta X^{(0)} = (1+\lambda_2)/(1-\lambda_2)$. 
Because of this quadrature mixing, a non-zero $\theta_{\rm rms}$ sets a limit to the maximum amount of squeezing that can be used.  By minimizing $(\Delta P)^2$ with respect to $\lambda_2$, the optimal parametric drive (i.e.~squeezing strength) is determined by: 
\begin{align}
    \frac{(1-\lambda_2)^4}{(1+\lambda_2)^4} = \theta_{\rm rms}^2.
\end{align}
Applying more squeezing than this limit increases $(\Delta P)^2$. 
To be safe, we want to err on the side of using less than the optimal amount of squeezing.  We thus use the following (heuristic) constraint (obtained by bounding the denominator on the RHS above by its maximum value):
\begin{align}
    (1-\lambda_2)^4 \gtrsim 16\theta_{\rm rms}^2,
\end{align}
The mean rotation angle $\theta_{\rm rms}$ is set by the RMS-averaged accumulated phase space rotation over the lifetime of the spectator mode $1/\kappa_{\rm c}$:
\begin{align}
    \theta_{\rm rms}^2 = \overline{\left[\phi(t)\right]^2};\qquad \phi(t)=\int_{0}^{1/\kappa_{\rm c}} dt \detr\xi_{\rm s}(t).
\end{align}
For Gaussian noise the limit on squeezing is thus
\begin{align}
    (1-\lambda_2)^4 \gtrsim 16 \detr^2 \int \frac{d\omega}{2\pi} \frac{S[\omega]}{\omega^2} \sin^2(\omega/2\kappa_{\rm c}). \label{eq:app-squeezing-limit}
\end{align}

To check whether this condition is indeed sufficient, we consider the limit of quasistatic noise, where we can perform exact calculations.  In this case, $\xi_{\rm s}(t) = \xi_{\rm s}$, a Gaussian random variable with variance $\sigma^2$.  The Gaussian parameter equations of motion are exactly solvable.
It can be shown that the most stringent limit on squeezing is set by the increased measurement imprecision noise due to amplitude quadrature noise being rotated into the squeezed phase quadrature.
This causes an increase in the long-time measurement imprecision noise dephasing $\LamImp(t\to\infty)$.
For each noise realization $\xi_{\rm s}$, $\LamImp(t\to\infty)$ is (up to a constant)
\begin{align}
    \LamImp(t\to\infty) = \frac{2\gff t}{(1+\lambda_2)^2}\left[ (1-\lambda_2)^2 + \frac{16\lambda_{2}\detr^{2}\xi_{\rm s}^{2}}{(1-\lambda_{2})^{2}\kappa_{\rm c}^{2}} \right],
\end{align}
where the first term in brackets is the dephasing due to the fluctuations in the squeezed phase quadrature, and the second term is the contribution due to the small rotation $\theta \propto \xi_{\rm s}/\kappa_{\rm c}$ of the amplitude quadrature into the phase quadrature.

We can now use this expression to determine when it is permissible to neglect the last term, i.e., the unwanted rotated noise from the amplified quadrature is valid.  We immediately obtain
\begin{align}
    (1-\lambda_2)^4 \gtrsim 16 \detr^2\xi_{\rm s}^2/\kappa_{\rm c}^2.
\end{align}
Taking the stochastic average of this condition over the classical noise then yields the quasistatic noise limit of our more general Eq.~\eqref{eq:app-squeezing-limit} derived above: 
\begin{align}
    (1-\lambda_2)^4 \gtrsim 16 \detr^2\sigma^2/\kappa_{\rm c}^2.
\end{align}


\subsection{Parameter estimation in a photonic state}
\label{app:parameter-estimation}

Here we briefly review the problem of estimating the small rotation angle of a squeezed displaced state and show that the estimation error exhibits Heisenberg scaling when the coherent state is correctly squeezed.
We start with a coherent state and show that the estimation error is bounded by the standard quantum limit in total photon number, then we consider a squeezed state and show an improved scaling of the estimation error with total photon number.

Suppose we have a coherent state $|\psi_0\rangle = \hat{D}(\alpha)|0\rangle$, where the displacement operator is $\hat{D}(\alpha) = \exp[\alpha\hat{a}^\dagger-\alpha^*\hat{a}]$.
Without loss of generality, we take the displacement to be real $\alpha >0$ which defines the amplitude quadrature [$\hat{X} = (\hat{a}^\dagger+\hat{a})/2$]. 
This state is rotated in phase space by a small angle $\theta\ll1$:
\begin{align}
    |\psi_\theta\rangle = {\rm e}^{{\rm i}\theta\hat{a}^\dagger\hat{a}}|\psi_0\rangle.
    \label{eq:rot-coh-state}
\end{align}
The problem is to estimate this angle given $\nbar = \alpha^2$ photons in the state $|\psi_\theta\rangle$. 
For small $\theta\ll1$, a measurement of the phase quadrature [$\hat{P} = {\rm i}(\hat{a}^\dagger-\hat{a})/2$] yields the estimate
\begin{align}
    \langle\hat{P}\rangle = \alpha\sin\theta \approx \alpha\theta = \sqrt{\nbar}\theta.
    \label{eq:param-estimate}
\end{align}
The uncertainty $\Delta P$ associated with this measurement depends on the variance of the phase quadrature:
\begin{align}
    (\Delta P)^2 =  \sqrt{\langle\hat{P}^2\rangle - \langle\hat{P}\rangle^2} = \frac{1}{4}.
\end{align}
The estimation error, $\Delta \theta = \Delta P/(|\partial \langle\hat{P}\rangle/\partial\theta|)$, is thus
\begin{align}
    \Delta\theta = \frac{1}{2\sqrt{\nbar}}
\end{align}
in terms of the total number of photons $\nbar$ in the state $|\psi_\theta\rangle$.
This is the standard quantum limit scaling of the estimation error with $\nbar$ measurement photons (i.e., $\nbar$ photons in the state $|\psi_\theta\rangle$) \cite{giovannetti_QuantumEnhanced_2004}. 

Now we turn to the case where the state is squeezed. 
The initial state is $|\psi_0\rangle = \hat{D}(\alpha)\hat{S}(-r)|0\rangle$, where the squeeze operator is $\hat{S}(\xi) = \exp[(\xi^*\hat{a}^2 - \xi\hat{a}^{\dagger2})/2]$ for $\xi=r{\rm e}^{{\rm i}\phi}$.
The phase $\phi=\pi$ is chosen to reduce the variance of the phase quadrature. 
Again the state is rotated in phase space by an angle $\theta\ll1$ (cf. Eq.~\eqref{eq:rot-coh-state}) and again the angle estimate is given by the measurement of the phase quadrature $\langle\hat{P}\rangle = \alpha\sin\theta \approx \alpha\theta$. 
The uncertainty $\Delta P$ associated with this measurement is now
\begin{align}
    (\Delta P)^2 \approx \frac{1}{4}{\rm e}^{-2r},
\end{align}
assuming a small rotation angle $\theta\ll {\rm e}^{-2r}$.
The estimation error is thus
\begin{align}
    \Delta \theta = \frac{1}{2\alpha}{\rm e}^{-r}.
\end{align}
The total number of photons in the squeezed coherent state is $\nbar$ = $\nbard$ + $\nbarsq$ where $\nbard = |\langle\hat{a}\rangle|^2 = \alpha^2$ is the displacement photon number, and $\nbarsq = \langle\hat{a}^\dagger\hat{a}\rangle - |\langle\hat{a}\rangle|^2 = \sinh^2 r$ is the squeezing photon number. 
For $r\gg1$ we have ${\rm e}^r = 2\sqrt{\nbarsq}$, thus the estimation error is $\Delta \theta = 1/4\sqrt{\nbard\nbarsq}$. 
Minimizing over the partition of the fixed total photon number $\nbar$ yields $\nbard = \nbarsq = \nbar/2$. 
Therefore, in terms of $\nbar$, the estimation error is
\begin{align}
    \Delta \theta = \frac{1}{\nbar},
\end{align}
which is Heisenberg-limited scaling of the estimation error with $\nbar$ measurement photons \cite{giovannetti_QuantumEnhanced_2004}.


\subsection{Autonomous spectator}
\label{app:autonomous}

As discussed in the main text, Eq.~\eqref{eq:qme} is not limited to describing the physical setup of homodyne measurement and feedforward.
As shown in Ref.~\cite{metelmann_Nonreciprocal_2017}, measurement and feedforward can be implemented autonomously using nonreciprocal interactions generated via engineered dissipation. 
Here we give a basic overview of how such an autonomous spectator system can be engineered.

To implement an autonomous spectator mode, nothing must change in Eq.~\eqref{eq:Hs} for $\hat{H}_{{\rm s},[\xi]}(t)$ nor in Eq.~\eqref{eq:Hq} for $\hat{H}_{{\rm q},[\xi]}(t)$.
Only the physical system which implements the coherent interaction $\hat{H}_{\rm int}$, Eq.~\eqref{eq:Hint}, and the dissipation $\hat{L}$, Eq.~\eqref{eq:L-diss}, must change.
Namely, $\hat{H}_{\rm int}$, must actually be implemented as written, and $\hat{L}$, must be engineered by coupling the qubit and spectator to a reservoir \cite{metelmann_Nonreciprocal_2015,metelmann_Nonreciprocal_2017}.
Recall that in the measurement and feedforward picture, the combination of dissipation and coherent interaction is a description of the homodyne detection and feedforward of the measurement record; no $\hat{H}_{\rm int}$ between the spectator and qubit is actually directly implemented.
These changes to the physical setup are shown in Fig.~\ref{fig:autonomous}.

A crucial element of the autonomous spectator implementation is the longitudinal coupling to the qubit required by $\hat{H}_{\rm int}$ (cf.~Eq.~\eqref{eq:Hint}), i.e., an interaction of the form
\begin{align}
    \hat{H}_{\rm long} = g \left({\rm e}^{{\rm i}\theta}\hat{d} + {\rm e}^{-{\rm i}\theta}\hat{d}^\dagger\right) \hat{\sigma}_{\rm z},
\end{align}
where $\hat{a}$ is a photonic mode and $\theta$ picks out a quadrature of that mode.
E.g., in the context of superconducting circuits, such couplings have been studied theoretically \cite{kerman_Quantum_2013,billangeon_CircuitQEDbased_2015,didier_Fast_2015} and demonstrated experimentally for linear resonators \cite{eichler_Realizing_2018}.
Longitudinal couplings have also been experimentally realized by dispersively coupling a cavity to a qubit and displacing the linear cavity \cite{touzard_Gated_2019}.

To engineer the collective dissipation generated by $\hat{L}$ (cf.~Eq.~\eqref{eq:L-diss}, we follow  Refs.~\cite{metelmann_Nonreciprocal_2015,metelmann_Nonreciprocal_2017} and introduce an auxiliary mode $\hat{b}$ with a damping rate $\gamma_{\rm b}$ to which we couple both the spectator and the qubit:
\begin{align}
    \hat{H}_{\rm aux} = \frac{1}{2}\sqrt{\gamma_{\rm b}\kappa_{\rm c}}\left(\hat{d}^\dagger\hat{b} + {\rm h.c.} \right) + \frac{1}{2}\sqrt{\gamma_{\rm b}\gff}\left(\hat{b} + \hat{b}^\dagger\right)\hat{\sigma}_{\rm z}.
\end{align}
By heavily damping this mode, $\gamma_{\rm b} \gg \kappa_{\rm c},\gff$, we can adiabatically eliminate it from the dynamics, thus deriving a dissipative interaction between the spectator and qubit.

The auxiliary Hamiltonian can be rewritten as
\begin{align}
    \hat{H}_{\rm aux} = \frac{1}{2}\sqrt{\gamma_{\rm b}} \left( \hat{b}^\dagger \hat{L} + {\rm h.c.} \right),
\end{align}
in terms of the collective jump operator $\hat{L}$.
Thus in the limit $\gamma_{\rm b} \gg \kappa_{\rm c}\gff$ we arrive at the dissipative interaction
\begin{align}
    \mathcal{D}[\hat{L}]\hat{\rho},
\end{align}
for the reduced spectator-qubit density matrix $\hat{\rho} = {\rm tr}_{\rm b}\{ \hat{\rho}_{\rm tot} \}$ (i.e., after tracing out the auxiliary mode $\hat{b}$).



\section*{Data Availability}

All data needed to evaluate the conclusions in the paper are present in the paper and/or appendices.


\section*{Acknowledgements}

This work was supported by the Army Research Office under Grant No. W911NF-19-1-0380. AC acknowledges support from the Simons Foundation, through a Simons Investigator award (Grant No. 669487, A. A. C.)


\section*{Competing Interests}

The Authors declare no Competing Financial or Non-Financial Interests.


\section*{Author Contributions}

A.L. and A.A.C. conceived the ideas and developed the theory. A.L. performed all calculations and numerical simulations and wrote the manuscript. A.L. and A.A.C. revised and edited the manuscript.

\bibliography{spectator-mode-qubit-dephasing}

\newpage


\appendix

\section{Feedforward delay.} 
\label{app:feedforward-delay}
Delay in the measurement and feedforward will degrade the noise mitigation performance of the spectator.
To include delay in the measurement and feedforward, we use the fact that the measurement is classical and nonreciprocal.
The classicality implies that the qubit and spectator never become entangled; thus from the perspective of the qubit, the spectator mode is another classical noise source that happens to be correlated with $\xi_{\rm q}(t)$.
The nonreciprocity implies that there is no feedback from the qubit to the spectator; thus there is no way for the qubit to learn about the feedforward delay.
Together these imply that from the perspective of the qubit, a delay in the measurement and feedforward, is equivalent to instantaneous measurement and feedforward of a delayed noise signal.
We therefore model the feedforward delay as a detection delay: at time $t$, the spectator detects delayed noise $\xi_{\rm s}(t-\tau_{\rm d})$ instead of the instantaneous noise $\xi_{\rm s}(t)$.
The delayed signal is then instantaneously fed forward to the qubit.

Due to the delay, the spectator transduction factor $\as$ picks up a phase factor 
\begin{align}
    \as \mapsto {\rm e}^{-{\rm i}\omega \tau_{\rm d}} \as
\end{align}
which reduces the spectator mode's ability to mitigate noise within its detection bandwidth.
In the long-time limit, however, the qubit remains sensitive only to the zero-frequency noise, thus the long-time dephasing rate vanishes for $\as=1$.

The effects of delay are best illustrated with an example. 
We consider a white noise spectral density $S[\omega]=S_0$ and a delay $\tau_{\rm d}$.
The decoherence function is
\begin{align}
    \chi&(t) = \LamImp(t) \label{eq:qubit-decoherence-delay} \\
    &+\begin{cases}
    S_0 t + \frac{S_0}{{\kappa_\phi}} \left( {\rm e}^{-{\kappa_\phi}t/2} - 1 \right) & t<\tau_{\rm d} \\
    S_0 \tau_{\rm d} + \frac{S_0}{{\kappa_\phi}} \left( 1 - 2{\rm e}^{-{\kappa_\phi}(t - \tau_{\rm d})/2} + {\rm e}^{-{\kappa_\phi}t/2}  \right) & t > \tau_{\rm d}\nonumber
    \end{cases}
\end{align}
where $\LamImp(t)$ is still given by Eq.~(26) of the main text. 
For pre-delay times $t<\tau_{\rm d}$ the spectator feedforward noise is uncorrelated with the direct noise on the qubit, leading to the initial dephasing at twice the bare rate, $S_0 t$.
There is also an exponential decay to a constant due to the Lorentzian spectral density of the noise fed forward from the spectator (bandwidth $\kappa_\phi$).
In the long-time limit the dephasing due to $\xi(t)$ is constant in time:
\begin{align}
    \chi(t\to\infty) = \LamImp(t) + S_0 \tau_{\rm d} + \frac{S_0}{{\kappa_\phi}}
\end{align}
where the constants are the delay-dependent dephasing ($S_0 \tau_{\rm d}$) and the zero-delay initial dephasing ($\chi_{\rm init}(\infty) = S_0/{\kappa_\phi}$, cf.~Eq.~(37) of the main text).

\begin{figure}[t]
     \centering
    \includegraphics[width=0.99\columnwidth]{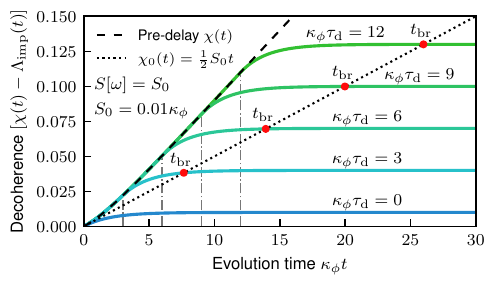}
     \caption{
        \textbf{Effects of feedforward delay on qubit decoherence.} 
    The qubit decoherence function (less $\LamImp(t)$) vs. time for various indicated delays $\tau_{\rm d}$ and for white noise $S[\omega] = S_0$ with $S_0 = 0.01\kappa_\phi$. 
    The dotted line is the bare qubit decoherence function and its intersection with the delay curves denotes the minimum break-even time $t_{\rm br}$ for that delay $\tau_{\rm d}$. 
    The dashed ``pre-delay $\chi(t)$'' curve is given by Eq.~\eqref{eq:qubit-decoherence-delay} evaluated as though $t<\tau_{\rm d}$ for all times, and the vertical dot-dashed lines indicate the delay time $\kappa_\phi\tau_{\rm d}$ for each delay curve -- this is the time when the spectator begins mitigating noise.}
     \label{fig:feedforward-delay}
 \end{figure}

The qubit decoherence function with feedforward delay is shown in Fig.~\ref{fig:feedforward-delay} for white noise. 
The bare dephasing of the qubit $\chi(t)=S_0 t/2$ intersects each delay dephasing function at the minimum possible the break-even time $t_{\rm br}$: the time at which the spectator system improves over the bare decoherence assuming negligible $\LamImp(t_{\rm br}) \ll S_0 t_{\rm br}$.
For $\tau_{\rm d}\gtrsim 1/{\kappa_\phi}$, the minimum break-even time is $t_{\rm br} \approx 2(\tau_{\rm d} + 1/{\kappa_\phi})$, and for $\tau_{\rm d} \ll 1/{\kappa_\phi}$ the minimum break-even time vanishes as $t_{\rm br} = (2+\sqrt{2})\tau_{\rm d}$.

\section{Internal loss and optimal squeezing.}
\label{app:internal-loss}

Throughout the main text, we assume that the coupling rate of the spectator to the output waveguide is the dominant source of damping and any internal loss is extremely weak: $\kappa_{\rm i}\ll\kappa_{\rm c}$.
Internal loss is experimentally unavoidable, however, and it effectively acts as a second pathway for the feedforward signal to leak out of the spectator.
This additional signal sink is most clearly an issue when the spectator is strongly squeezed as it limits how much the vacuum noise in the output field can be reduced.

 \begin{figure}[t]
     \centering
    \includegraphics[width=0.999\columnwidth]{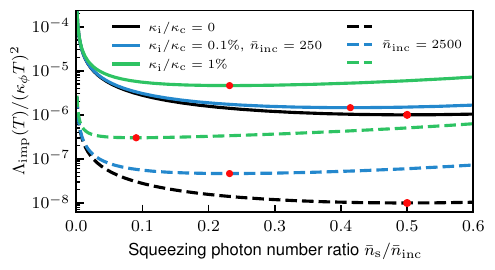}
     \caption{
     \textbf{Effects of spectator mode internal loss on the optimal ratio of temporal mode squeezing photons.}
     $\LamImp(T)/(\kappa_\phi T)^2$ (cf. Eq.~\eqref{eq:lambda-add-internal-loss}) vs. squeezing photon number ratio $\nbarsq/\nbartot$ in the presence of indicated degrees of internal loss $\kappa_{\rm i}/\kappa_{\rm c}$, for a fixed long evolution time $T\gg1/\kappa_\phi$ and $g=1$. 
     We show $\nbartot=250$ in solid lines and $\nbartot=2500$ in dashed lines; the colors denote various $\kappa_{\rm i}/\kappa_{\rm c}$.
     The minimum of each curve is marked by a red dot and denotes the optimal fraction of the total photon number which should be used in the squeezing to minimize the long-time dephasing rate.
     Note that for a given $\kappa_{\rm i}/\kappa_{\rm c}$, the optimal squeezing photon number ratio decreases for increasing $\nbartot$; this is the breakdown of Heisenberg-limited scaling with internal loss.
     }
     \label{fig:internal-loss}
 \end{figure} 

We let the mode couple to a zero temperature internal loss bath (although this can be generalized to finite temperature in a straightforward manner) with rate $\kappa_{\rm i}$.
The total loss rate is
\begin{align}
    \kappa_{\rm tot} = \kappa_{\rm c} + \kappa_{\rm i}
\end{align}
The drive strengths must be increased in proportion to the damping rate; the overall rate scale $\kappa_{\rm c}$ in the drive Hamiltonian (cf.~Eq.~(5) of the main text) is replaced by $\kappa_{\rm c} \mapsto \kappa_{\rm tot}$.
The ideal spectator transduction factor is now
\begin{align}
    \as^{\rm ideal} = \frac{\kappa_{\rm tot}}{\kappa_{\rm c}} \geq 1,
\end{align}
and all of the properties of the spectator at ideal transduction strength otherwise hold.

The internal loss effects a qualitative change to the measurement imprecision noise dephasing $\LamImp$:
\begin{align}
    \LamImp(t)&=\frac{\alpha_{\rm s}^{2}}{32\detr^{2}\nbardi} \frac{\kappa_{\rm c}}{\kappa_{\mathrm{tot}}} \left[ \frac{(1-\lambda_{2})^{2}}{1+\lambda_{2}}+\frac{\kappa_{\rm i}}{\kappa_{\rm c}}(1+\lambda_{2})\right]\kappa_{\phi}t.
\end{align}
Here we neglect the exponential decay to a constant. These terms each receive a $(\kappa_{\rm c}/\kappa_{\rm tot})^2$ prefactor, and the exponential gets the replacement $\kappa_\phi\mapsto\kappa_{{\rm tot},\phi}$.
The linear-in-$t$ term is no longer simply proportional to $(1-\lambda_2)^2$.
The term $\propto \kappa_{\rm i}$ is caused by the splitting of the mode's output field between the internal loss and the waveguide, which limits the degree of squeezing in the waveguide.

In terms of the number of photons in the temporal mode $\hat{A}$ (cf.~Eq.~(43) of the main text), and for $\nbartot,\nbard,\nbarsq\gg1$, the long-time measurement imprecision noise dephasing is now given by
\begin{align}
    \LamImp(T) = \frac{(\kappa_\phi T)^2}{64 \detr^2}\frac{\kappa_{\rm tot}}{\kappa_{\rm c}}\left[ \frac{1}{\nbard \nbarsq} + \frac{2\kappa_{\rm i}/\kappa_{\rm c}}{\nbard} \right]
    \label{eq:lambda-add-internal-loss}
\end{align}
where we have let $\as = \kappa_{\rm tot}/\kappa_{\rm c}$, its ideal value. 
For a fixed $\nbartot$ in the temporal mode, the optimal choice of $\nbard$ and $\nbarsq$ is no longer $\nbard = \nbarsq = \nbartot/2$, but is now dependent on the internal loss ratio $\kappa_{\rm i}/\kappa_{\rm c}$, as we show in Fig.~\ref{fig:internal-loss}.
The optimal choice rapidly approaches $\nbard = \nbartot$ when the product $\nbartot(\kappa_{\rm i}/\kappa_{\rm c})\gg1$.
Furthermore the Heisenberg-limited scaling $\LamImp(T)\propto 1/\nbartot^2$ rapidly breaks down as the internal loss term in Eq.~\eqref{eq:lambda-add-internal-loss} becomes dominant with increasing $\nbartot$ because it is $\propto \nbard \sim 1/\nbartot$ instead of $\sim 1/\nbartot^2$.


\end{document}